\documentclass[a4paper]{article}%,draft
\usepackage{amsmath}
\usepackage{amsthm}

\usepackage[english]{babel}
\usepackage[utf8]{inputenc}
\usepackage{graphicx}
\usepackage[colorinlistoftodos]{todonotes}
\usepackage{pdfpages}
\usepackage[margin=1.0in]{geometry}
\usepackage{caption}
\usepackage{subcaption}
\usepackage{multirow}
\usepackage{tabularx}
\usepackage{float}
\usepackage{MnSymbol}
\usepackage{hyperref}
\usepackage{mathtools}
\usepackage{amsfonts}
\usepackage[linesnumbered,ruled,vlined]{algorithm2e}

\usepackage{listings}
\renewcommand{\vec}[1]{\boldsymbol{#1}}

\newcommand{\dd}[0]{\mathrm{d}}
\newcommand{\vc}[1]{\boldsymbol{#1}}

\newcolumntype{Y}{>{\centering\arraybackslash}X}
\newcolumntype{s}{>{\hsize=1.1cm}X}

\SetCommentSty{mycommfont}

\title{On the periodicity of cardiovascular fluid dynamics simulations}
\author{Martin~R.~Pfaller, Jonathan~Pham, Nathan~M.~Wilson, David~W.~Parker, Alison~L.~Marsden}
\begin{document}
\maketitle

%\tableofcontents

\begin{abstract}

Three-dimensional cardiovascular fluid dynamics simulations typically require computation of several cardiac cycles before they reach a periodic solution, rendering them computationally expensive. Furthermore, there is currently no standardized method to determined whether a simulation has yet reached that periodic state. In this work, we propose use of the asymptotic error measure to quantify the difference between simulation results and their ideal periodic state using lumped-parameter modeling. We further show that initial conditions are crucial in reducing computational time and develop an automated framework to generate appropriate initial conditions from a one-dimensional model of blood flow. We demonstrate the performance of our initialization method using six patient-specific models from the Vascular Model Repository. In our examples, our initialization protocol achieves periodic convergence within one or two cardiac cycles, leading to a significant reduction in computational cost compared to standard methods. All computational tools used in this work are implemented in the open-source software platform SimVascular. Automatically generated initial conditions have the potential to significantly reduce computation time in cardiovascular fluid dynamics simulations.

\end{abstract}

\section{Introduction}
Three-dimensional (3D) blood flow simulations are commonly coupled with zero-dimensional (0D) lumped parameter models, representing the downstream vasculature at the model's boundary \cite{kung13,seo19,fleeter20,arthurs20}. These lumped parameter models are analogous to an electric circuit, with resistors and capacitors modeling the viscosity of the blood and the elasticity of the vessel wall, respectively. A popular choice is the three-element Windkessel model, also known as the RCR model \cite{vignonclementel10}. The Windkessel consists of a proximal resistance in series with a parallel distal resistance and capacitance (Figure~\ref{fig_flowchart}).

The RCR boundary condition, like many other outflow boundary conditions, contains a capacitor that stores blood volume. While these capacitors are ``charging", it typically takes several cardiac cycles with a periodic pulsatile inflow to reach a periodic state. A periodic state is here defined as two consecutive cardiac cycles yielding  results for pressure and flow rate that agree within a given tolerance. Mathematically, this corresponds to the limit cycle of the model. It is essential to extract simulation results only when they have reached this periodic state, e.g., when comparing them to \textit{in vivo} measurements. However, there are currently no clear guidelines on how to determine whether this periodic state has been achieved.

Unfortunately, running several cardiac cycles of a 3D simulation is computationally expensive, typically requiring a high-performance computer. The computation time scales linearly with the number of cardiac cycles since they cannot be run in parallel. A common practice is initializing a simulation with results from a steady-state solution \cite{vignonclementel10}. Here, a computationally inexpensive simulation with constant inflow is computed first. Its solution is then used as an initial condition for the simulation with a periodic pulsatile inflow condition.  However, as we will show in this work, a simulation initialized with a steady-state solution often still requires several cardiac cycles to reach a reasonably periodic state.

Our goal is twofold. First, we introduce metrics and tools to quantify whether a simulation has yet reached a periodic state within a given tolerance. Second, we shorten the computation time of the computationally expensive 3D simulation by leveraging reduced-order modeling. It was previously shown that one-dimensional (1D) models of cardiovascular fluid dynamics could accurately approximate integral quantities of 3D solutions such as velocity and pressure at the outlets \cite{wan02,moore04,grinberg10,reymond12,xiao13,pant14,bertaglia20}. In this work, we propose a novel method to initialize a 3D simulation using the results from an inexpensive 1D simulation. This framework is fully automated and requires no user interaction. %Toward this goal, we present a framework that automatically generates a 1D model from an existing 3D model. We start with theoretical considerations on how to determine if the solution has yet arrived at a periodic state. We then propose a technique to project the pressure and velocity from a  1D solution to the 3D mesh. 

\section{Methods}
We begin by revisiting the governing equations of 3D, 1D, and 0D fluid dynamics in Sections~\ref{sec_mod_3d}, \ref{sec_mod_1d}, and \ref{sec_mod_0d}. Following theoretical considerations of the convergence of lumped parameter models, we define suitable error metrics to determine the difference between a 3D simulation and its periodic state in Section~\ref{sec_rcrc}. In Section~\ref{sec_3D_periodicity}, we introduce a tool to determine whether a simulation has yet reached its periodic state. Finally, we develop a method to initialize a 3D simulation from a 1D simulation in Section~\ref{sec_ini} in order to jumpstart initialization and reduce computational cost.

\subsection{3D flow physics \label{sec_mod_3d}}
The dynamics of blood flow in the cardiovascular system is mathematically governed by the incompressible 3D Navier-Stokes equations, 
\begin{align}
\rho \left(\dot{\vc{v}} + \vc{v} \cdot \nabla \vc{v} \right) &= - \nabla P + \mu\nabla^{2}\vc{v} + \vc{f},& \text{in~} & \Omega,
\label{NS}\\
\nabla \cdot \vc{v} &= 0,  & \text{in~} & \Omega.
\label{continuity}
\end{align}
The first equation in this system represents conservation of momentum for a Newtonain, incompressible fluid, where $\vc{v}$ is the velocity of the blood flow, $P$ is pressure, $\vc{f}$ is a body force, $\rho$ is the density of the blood, and $\mu$ is the dynamic viscosity. The second equation in this system represents conservation of mass. %The derivation of the Navier-Stokes equations can be found in many references, including classical texts such as (insert references, i.e. Batchelor, etc. ). 
%
% a differential form of Newton's second law, applied to fluid parcels. 
%
% The Navier-Stokes equations \cite{quarteroni16}, 
%
% \begin{a%lign}
% \rho \left(\dot{\vc{v}} + \vc{v} \cdot \nabla \vc{v} \right) &= - \nabla P + \mu\nabla^{2}\vc{v} + \vc{b},& \text{in~} & \Omega,
% \label{NS}\\
% \nabla \cdot \vc{v} &= 0,  & \text{in~} & \Omega,
% \label{continuity}
% \end{align}
%
% where $\vc{v}$ is velocity, $P$ is pressure, $\vc{b}$ is body force, $\rho$ is density, and $\mu$ is dynamic viscosity, govern many fluid phenomena, including dynamical, three-dimensional blood flow in the cardiovascular system. The first equation in this system represents the differential form of Newton's second law, applied to fluid parcels. The left-hand side of this equation captures the acceleration of the fluid elements, while the right-hand side represents the pressure forces, viscous forces, and body forces acting on the fluid. The second equation represents mass conservation in the system. The derivation of the Navier-Stokes equations can be found in many references, including classical texts such as (insert references, i.e. Batchelor, etc. ). 
%
In the computational cardiovascular modeling and simulation context, we typically numerically solve the 3D Navier-Stokes equations in patient-specific models of vascular anatomies to simulate hemodynamics. Simulation results are used to elucidate the relationship between cardiovascular diseases and fluid mechanics, for personalized treatment planning, and to aid the development of novel biomedical technologies. The initial conditions for velocity and pressure are
\begin{align}
\vc{v}(\vc{x}, t = 0) = \vc{v}_0(\vc{x}), \quad P(\vc{x}, t = 0) = P_0(\vc{x}), %\label{oneD_ICs}
\end{align}
Boundary conditions that model the portion of the cardiovascular system not captured by the anatomical model must be provided as well. A flow rate, $Q$, is commonly prescribed at the inlet surfaces of the 3D vascular model, where the flow rate is computed via integration of the normal velocity over each inlet surface, %velocity component normal to the surface of interest,
\begin{align}
Q(\vc{x}, t) &= \int_\Gamma \vc{v} \cdot \vc{n} \,d\Gamma = Q_{in}(t),  & \text{in~} & \Gamma_{on}. \label{threeD_in_bc}
\end{align}
On the other hand, lumped parameter models, also known as 0D models, are commonly used as boundary conditions at the outlets of the model. These lumped parameter models usually relate the pressure to the flow rate via parametric differential-algebraic equations, 
% On the other hand, pressures, $P$, are commonly used as boundary conditions at the outlets of the model. Here, pressures outlet boundary conditions are often expressed as parametric functions of flow rate via ordinary differential equations,
%
\begin{align}
P(\vc{x}, t) &= f(\vc{x}, t, Q_{out}(t), \dot{Q}_{out}(t), \vc{\phi}),  & \text{on~} & \Gamma_{out}, \label{threeD_out_bc}
\end{align}
where $\vc{\phi}$ represents the set of variables parametrizing the differential equation. A discussion of 0D models and some commonly used outlet boundary conditions is provided in section 2.3. 

We generate 3D patient-specific models using SimVascular, an open-source, comprehensive modeling, and simulation software for vascular anatomies (\url{simvascular.org}) \cite{updegrove16}. The models are simulated in our open-source solver svSolver using the Finite Element Method (FEM) (\url{github.com/SimVascular/svSolver}). It uses liner P1-P1 elements with a streamline upwind Petrov-Galerkin and pressure-stabilizing Petrov-Galerkin formulation (SUPG/PSPG) \cite{franca92}. The pressure and momentum stabilization is detailed in \cite{taylor98,whiting01}. Furthermore, it uses a linear solver with specialized pre-conditioners tailored to handle large vascular resistances coupled at outflow boundaries \cite{esmailymoghadam13}. All 3D simulations in this work were run on Stanford's Sherlock supercomputing cluster using four 12-core Intel Xeon Gold 5118 CPUs.

%In SimVascular, the 3D Navier-Stokes equations, coupled to initial conditions and boundary conditions, are numerically solved using a variational multiscale formulation of the finite element method with an implicit generalized-alpha time-stepping scheme (insert reference to Ken jansen's generalized alpha paper).

% Despite the powerful, non-invasive advantage that 3D models play in cardiovascular scientific inquiry and engineering, 3D models are not widely used in clinical practice yet. The expensive 

% However, one of the significant impediments of 3D models if that they can take many hours to days to simulate, even on multi-processor supercomputers. 

%These 3D anatomical models are often coupled with boundary conditions governed by ordinary differential equations

\subsection{1D flow physics \label{sec_mod_1d}}

In contrast to 3D models, one-dimensional (1D) models have only a single spatial dimension, the axial dimension along the centerline of the vessel \cite{hughes73,wan02,steele03}. Due to this lack of 3D spatial information, 1D models are capable of simulating only bulk flow rate, and cross-sectionally averaged pressure at each centerline node of the 1D finite element model. We integrate incompressible 3D Navier-Stokes equations \eqref{NS} over the cross-section while assuming Newtonian fluid properties for the blood and an axisymmetric parabolic flow profile to obtain the governing equations for the 1D model. This process yields
\begin{align}
\dot{S} + \frac{\partial Q}{\partial z} &= 0,\\
%\label{1Dcontinuity}
\dot{Q} + \frac{4}{3} \frac{\partial}{\partial z} \frac{Q^2}{S} + \frac{S}{\rho}\frac{\partial P}{\partial z} &= Sf -8\pi\nu \frac{Q}{S}+\nu \frac{\partial^2 Q}{\partial z^2},
%\label{1Dmomentum}
\end{align}  
with flow rate $Q$, cross-sectional area $S$, pressure $P$, density $\rho$, body force $f$, and kinematic viscosity $\nu$. The coordinate $z$ represents the axial dimension of 1D model. To solve these equations, we also require a constitutive law to relate the pressure to the cross-sectional area. In this work, we use the constitutive relationship proposed by Olufsen \cite{olufsen99}, 
\begin{align}
 P(z,t)=P^0(z)+\frac{4}{3}\frac{Eh}{r^0(z)} \left( 1-\sqrt{\frac{S^0(z)}{S(z,t)}} \right), \quad
\frac{Eh}{r^0(z)}=k_1e^{k_2r^0(z)+k_3},
\end{align}
where $E$ is the Young's modulus of the blood vessel, $h$ is the wall thickness, $P_0$ is the reference pressure, $r_0$ is the reference radius, and $k_1$, $k_2$, and $k_3$ are empirically derived constants. % Furthermore, for models with multiple vascular branches, we enforce pressure continuity and conservation of mass at the bifurcations using Lagrange multipliers. 

%This formulation of the 1D governing equations enables the 1D model to capture deformations of the vessel walls, if desired. 
Furthermore, for models with multiple vascular branches, mass conservation is obeyed at the junction regions, and the pressure is assumed to be constant between the inlet and outlets of the junctions \cite{wan02}. 
%Although not taken into account in this work, the pressure drop across stenosis also can be modeled \cite{mirramezani19}. 
As with 3D models, we need initial conditions to initialize the simulation, 
\begin{align}
Q(z, t = 0) = Q_0(z), \quad P(z, t = 0) = P_0(z), \quad S(z, t = 0) = S_0(z), %\label{oneD_ICs}
\end{align}
as well as inlet and outlet boundary conditions, discussed in section 2.3, to represent the portion of the cardiovascular not reflected in our 1D model,
\begin{align}
% Q(z, t) &= Q_{in}(t), & \text{in~} & \Gamma_{in}, \label{oneD_in_bc} \\
% Q(z, t) = Q_{in}(t), \quad \text{in} \quad \Gamma_{in}, \label{oneD_in_bc} \\
Q(z, t) &= Q_{in}(t),  & \text{in~} & \Gamma_{in}, \label{oneD_in_bc} \\
% Q(z, t) &= Q_{out}(t), \text{or} P(z, t) &= P_{out}(t), \text{or} P(z, t) &= R*Q_{out}(t), & \text{in~} & \Gamma_{out}, \label{oneD_out_bc}
% Q(z, t) = Q_{out}(t), \quad \text{or} \quad P(z, t) = P_{out}(t), \quad \text{or} \quad P(z, t) = R*Q_{out}(t) \quad \text{in} \quad \Gamma_{out}, \label{oneD_out_bc}
% \quad P(z, t) = f(z, t, Q_{out}(t), \vc{\phi}), \quad \text{in} \quad \Gamma_{out}, \label{oneD_out_bc}
\quad P(z, t) &= f(z, t, Q_{out}(t), \dot{Q}_{out}(t), \vc{\phi}),  & \text{in~} & \Gamma_{in}, \label{oneD_out_bc}
\end{align}
We generate and simulate the 1D centerline and finite element models using SimVascular and VMTK \cite{antiga08}. We solve the set of differential equations using our open-source solver svOneDSolver (\url{github.com/SimVascular/svOneDSolver}). We employ a stabilized space-time finite element method based on the discontinuous Galerkin method in time \cite{wan02}. The spatial discretization employs continuous piecewise linear polynomials whereas we use a piecewise constant temporal discretization. For more background information, see \cite{brooks82,hughes86,hughes89}.

% In our simulation results presented below, we consider rigid wall behavior and obtain this effect by artificially applying a large Young's modulus for the blood vessel's material properties.

\subsection{0D flow physics \label{sec_mod_0d}}

The third model fidelity we consider in this work is the zero-dimensional (0D) model. Unlike 3D and 1D models, 0D models lack spatial information. However, as with 1D models, 0D models are capable of accurately simulating bulk flow rate and bulk pressure quantities in the cardiovascular system \cite{mirramezani20}, \cite{formaggia09}. These 0D models are composed of individual lumped-parameter elements that connect to form an entire complex lumped parameter network. There are many lumped parameter elements commonly used in the context of cardiovascular modeling and simulation. Some of these elements include resistors, capacitors, and inductors \cite{formaggia09}. Resistors model the viscous effects of blood flow, capacitors represent the elastic nature of blood vessels, while inductors capture the inertia of the blood flow. In lumped parameter networks that model the heart, diode elements are also employed. These diodes mimic the behavior of heart valves, where they allow flow to pass only when the valves are open \cite{kim09}, \cite{schiavazzi16b}. The flow rates and pressures in each of these elements are respectively governed by the following linear differential and algebraic equations,
\begin{align}
\Delta P = RQ, \quad Q = C\Delta \dot{P}, \quad \Delta P = L\dot{Q}, \quad Q = \frac{|Q| + Q}{2},
\end{align} 
where $R$ is the resistance, $C$ is the capacitance, $L$ is the inductance, and $\Delta P$ is the pressure drop across the element. Lumped parameter 0D models are also analogous to electrical circuits, where the flow rate and pressure are representative of current and voltage, respectively. In this work, we will focus our attention to just resistors and capacitors.

% Alison: if you include a heart model, there are other important circuit elements such as diodes.   I'd recommend editing this section to reflect these more general scenarios.

% MIGHT BE WORTH MENTIONING THAT 0D MODELS ARE ANALAGOUS TO ELECTRICAL CIRCUITS HERE (might be useful in later discussions, when we discuss how RCR BCs need time to charge and fill up, before we hit periodic convergence). 

Lumped parameter networks are commonly used in two contexts. First, 0D models can be used as surrogate models of entire vascular anatomies \cite{mirramezani20}. In this case, each blood vessel in the vascular system is represented by one or more lumped-parameter elements in the 0D model, where the value of each element is determined by the geometric and material properties of the blood and blood vessel. Second, 0D models can be used to represent boundary conditions in 3D and 1D models \cite{kim09}, \cite{vignon04}, \cite{esmailymoghadam13}, \cite{mirramezani19}, \cite{vignonclementel06}. In this context, each 0D element reflects a different downstream (or upstream) anatomical feature of the cardiovascular system. To employ the 0D models as boundary conditions, they must be numerically coupled to the 3D or 1D finite element models \cite{esmailymoghadam13}. The governing equations for these 0D models then can be numerically solved using traditional time-stepping schemes, such as an explicit fourth-order Runge-Kutta method.  % In this work, we couple our models using the approach presented in \cite{esmailymoghadam13}. %In this work, we employ 0D models as boundary conditions for the outlets of our 3D and 1D models. 

\begin{figure}[hbt!]
\centering
\includegraphics[width=1.\textwidth]{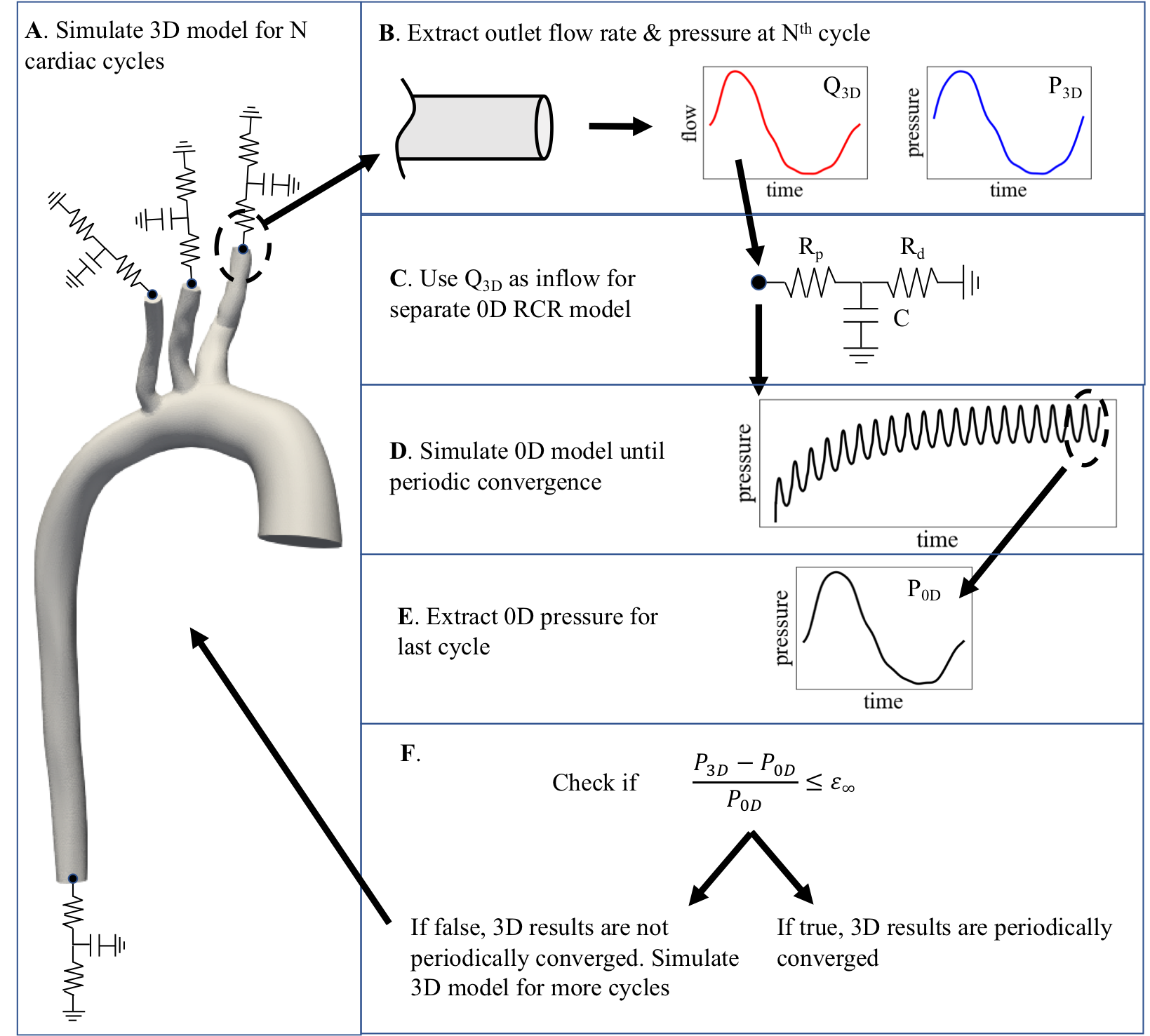}
\caption{Flowchart of method to check periodic state of 3D simulation results.\label{fig_flowchart}}
\end{figure}

Some of the 0D models most commonly employed as boundary conditions are the resistance model and the 3-element Windkessel model \cite{vignonclementel10,vignonclementel06}. The resistance model is composed of a single linear resistor element that captures the downstream resistance of the vascular network not portrayed in the 3D or 1D model. On the other hand, the 3-element Windkessel model, also known as the RCR model, as shown in Figure~\ref{fig_flowchart}, models the proximal resistance, distal resistance, and compliance of the downstream vasculature using two resistor elements and a capacitor. The RCR boundary condition is discussed further in the next section.

To simulate our 0D surrogate models, which represent the vessels using linear resistors, we solve the governing system of equations using an in-house, modular numerical solver that employs the implicit generalized-$\alpha$ scheme for time advancement \cite{jansen00}.  %Our numerical solver also includes a Newton-Raphson iterative solver, but in this work, we solely consider linear 0D models.

% The 0D models must be numerically coupled to the 3D finite element models \cite{ESMAILYMOGHADAM201363}. This coupling is already implemented in our 

% In order to use these 0D models as boundary conditions in our 3D and 1D finite element models, we must couple the 
% The 0D models are often solved numerically using traditional time-stepping schemes. In this work, we solve the 

% Alison: add reference to the 0D/3D coupling paper and mention the need for coupling numerically.   Also add a sentance on how we solve the 0D systems (numerically or analytically depending on the situation, often with an RK4 method, etc). 

% During the 0D simulation, the equations for each lumped parameter element are combined into a linear system of differential-algebraic equations, 

% %
% \begin{align}
% E\dot{y} + Fy + C = 0,
% %\label{0D_DAE}
% \end{align}  
% %

% where $y$ represents the vector of flow rates and pressures in our 0D model. In our work, we numerically solve this system for $y$ using the generalized-alpha method.

% $E$ is a matrix containing the capacitance and inductance quantities, $F$ is a matrix containing the resistance parameters, and $C$ is a vector containing the pressure drop indicated in the left-hand side of the resistor equation, 

\subsection{The RCR boundary condition \label{sec_rcrc}}
Thus far, we have introduced three different model fidelities commonly used in computational vascular modeling and simulation. Before any of the models and simulation results can be used in predictive and scientific applications, the quantities of interest, primarily the flow rates and pressures, must be simulated until they converge to a periodic state. Here, a simulated quantity of interest is considered to be periodic if its values between 2 adjacent periods are the same, within some defined tolerance. A period in the cardiovascular context is typically defined as a single cardiac cycle. %Mathematically, this definition is expressed as: ...
We expound on the concept of a periodic state in this section, using the RCR model to illustrate the primary concepts. 

The three parameters of the RCR boundary condition are commonly tuned to clinical measurements, i.e., phase-contrast magnetic resonance imaging and pressure measurements by solving an optimization problem to match minimal and maximal pressure over a cardiac cycle, measured flows, and flow distribution among different outlets \cite{spilker10}. A common strategy is to tune the total resistance and capacitance in the model to produce a physiologic pressure waveform, then distributed proportional to the vessel outlet areas \cite{zhou99}.

We begin by reviewing the response of a single RCR model to pulsatile inflow. The governing differential equation for the inlet pressure, $P$, of the 3-element Windkessel model, as a function of the inlet flow rate, $Q$, is
\begin{align}
\dot{P} + \frac{P}{\tau} = R_p\dot{Q} + \frac{1}{\tau}(R_p + R_d)Q, \quad \tau = R_dC,
%\label{0D_DAE}
\end{align}  
where $R_p$ is the proximal resistance, $C$ is the capacitance, and $R_d$ is the distal resistance. 
%
%Using the flow rate, $Q$, obtained from the 3D or 1D model as the input to equation 7, the pressure, $P$, at the inlet of the RCR boundary condition can be solved for. This $P$ is then used to update the 3D or 1D pressures and flow rates. The process repeats iteratively, using the 3D and 1D solvers described in sections 2.1 and 2.2, until the solution converges to the user-specified solver tolerances.
%
We obtain the semi-analytical solution \cite{vignon04} for this ordinary differential equation as
\begin{align}   
P(t) = \underbrace{[P(0) - R_p Q(0)] \, e^{-t/\tau}}_{\text{I}} + \underbrace{R_p Q(t)}_{\text{II}} + \underbrace{\int_0^t \frac{e^{-(t-\tilde{t})/\tau}}{C} Q(\tilde{t}) ~ \dd \tilde{t}}_{\text{III}}, \quad \tau = R_dC > 0,
\end{align}
which depends on the inflow $Q$ and the time constant $\tau$. We can identify three different terms in this equation: (I) exponential decay of the initial solution, (II) pressure drop at proximal resistance, (III) pressure drop at sub-circuit $R_dC$. Assuming a constant inflow $\bar{Q}>0$ for $t>0$, we obtain the pressure step response for the RCR boundary condition as
\begin{align}
P(t) = P_\infty + e^{-t/\tau} \, [P_0 - P_\infty], \quad P_0 = P(0), \quad \lim_{t\to\infty} P(t) = P_\infty = \bar{Q}(R_p + R_d),
\label{eq_step_constant}
\end{align}
starting at the initial pressure $P_0$ and exponentially approaching the asymptotic pressure $P_\infty$ for the limit $t\to\infty$. We now define the periodic inflow $Q(t)$ as
\begin{align}
Q(t + T) = Q(t), \quad \bar{Q} = \frac{1}{T} \int_0^T Q(t) ~ \dd t, %\quad t \in [0, T), 
\end{align}
where the period $T$ is the length of a cardiac cycle and $\bar{Q}$ the time-averaged mean flow. The mean pressure in the $n$-th cardiac cycle is denoted by
\begin{align}
\bar{P}_n = \frac{1}{T} \int_{nT}^{(n+1)T} P(t) ~ \dd t, \quad n \in \mathbb{N}^+_0.
\end{align}
With this notation, Equation~\eqref{eq_step_constant} can be reformulated for a non-constant, periodic pulsatile inflow as
\begin{align}
%P_n(t) = P_\infty(t) + e^{-(t+nT)/\tau} \, [P_0(t) - P_\infty(t)], \quad P_\infty(t) = \lim_{n\to\infty} P_n(t)
\bar{P}_n = \bar{P}_\infty + e^{-nT/\tau} \, [\bar{P}_0 - \bar{P}_\infty], \quad \bar{P}_\infty = \lim_{n\to\infty} \bar{P}_n,
\label{eq_step_puls}
\end{align}
starting at the initial mean pressure $\bar{P}_0$ and approaching the asymptotic mean pressure $\bar{P}_\infty$. Note that the asymptotic mean pressure $\bar{P}_\infty$ can in general not be determined analytically but depends on the function $Q(t)$. Notably, $\bar{P}_\infty$ is different from the asymptotic pressure $P_\infty$ that $P$ approaches for a solution with steady mean flow $\bar{Q}$.

In the remainder of this section, we will define an error metric to quantify the difference between the pressure $\bar{P}_n$ in the $n$-th cardiac cycle and the asymptotic pressure $\bar{P}_\infty$. For simplicity of notation, we show the following derivations for $\bar{P}_n < \bar{P}_\infty$, i.e. the pressure approaches the asymptotic pressure ``from below''. However, the conclusions hold for any choice of initial pressure. We define the asymptotic error $\epsilon_\infty$ as
\begin{align}
0 < \epsilon_\infty \leq \frac{\bar{P}_\infty - \bar{P}_n}{\bar{P}_\infty} = e^{-nT/\tau} \cdot \left( 1 - \frac{\bar{P}_0}{\bar{P}_\infty} \right).
\label{eq_pressure_error_tolerance}
\end{align}
Thus, we can calculate the number of cardiac cycles $n_\infty$ required for the RCR boundary condition to reach a periodic state within the tolerance $\epsilon_\infty$ as
\begin{align}
n_{\infty} \leq - \frac{\tau}{T} \cdot \ln{\frac{\epsilon_\infty}{1 - \bar{P}_0/\bar{P}_\infty}}, \quad n_\infty \in \mathbb{N}^+.
\label{eq_err_asym}
\end{align}
Several observations can be made from Equation~\eqref{eq_err_asym}. First, the number of cycles scales linearly with $\tau/T$, the ratio of the length of the time constant $\tau$ to the length of one cardiac cycle. A two-fold increase in the time constant doubles the number of cardiac cycles required to reach the same tolerance $\epsilon_\infty$. Second, the closer the initial pressure $\bar{P}_0$ to the asymptotic pressure $\bar{P}_\infty$ the fewer cardiac cycles $n_{\infty}$ are required to achieve periodic convergence with an error $\epsilon_\infty$. The error $\epsilon_{\infty}$ and the number of cardiac cycles $n_{\infty}$ can in general not be determined \textit{a priori} since the constant $\bar{P}_\infty$ cannot be evaluated analytically. Only in the special case of starting from zero initial conditions, i.e. $P_0=0$, the number of cardiac cycles to periodic convergence can be directly given as
\begin{align}
n_{\infty} \leq - \frac{\tau}{T} \cdot \ln \epsilon_\infty.
\label{eq_err_asym_zero}
\end{align}
We can easily compute the difference $\bar{P}_n - \bar{P}_{n-1}$ between two cardiac cycles numerically without the solution having reached a periodic state. Using this difference, we define the cyclic error $\epsilon_n$ between two consecutive cardiac cycles as
\begin{align}
%\epsilon_T &= \frac{P(t) - P(t-T)}{P_\infty} = e^{-t/\tau} \cdot \left( 1 - \frac{P_0}{P_\infty} \right) \cdot [e^{T/\tau} - 1 ]
0 < \epsilon_n & \leq \frac{\bar{P}_{n} - \bar{P}_{n-1}}{\bar{P}_\infty} = e^{-nT/\bar{\tau}} \cdot \left( 1 - \frac{\bar{P}_0}{\bar{P}_\infty} \right) \cdot [e^{T/\tau} - 1 ], \quad n \geq 2.
\label{eq_err_cyc}
\end{align}
We then define the ratio $\alpha$ between asymptotic error $\epsilon_{\infty}$ and cyclic error $\epsilon_n$ as
\begin{align}
\alpha = \frac{\epsilon_\infty}{\epsilon_n} = \frac{\bar{P}_\infty - \bar{P}_n}{\bar{P}_n - \bar{P}_{n-1}} = \frac{1}{e^{T/\tau} - 1}, \quad \leadsto ~ \epsilon_\infty > \epsilon_n \text{~, for~} \frac{\tau}{T} > \frac{1}{\ln2} \approx 1.44
\label{eq_alpha}
\end{align}
Equation \eqref{eq_alpha} shows that in general $\epsilon_n \neq \epsilon_\infty$. While error metrics like $\epsilon_n$ are commonly used in practice to determine whether a simulation as  reached a periodic state within a given tolerance, it is not an adequate metric to do so. For time constants $\tau$ larger than 1.44 times the length of the cardiac cycle $T$, the cyclic error $\epsilon_n$ will underestimate the asymptotic error $\epsilon_\infty$. In simulations with a large time constant $\tau \gg T$ the error $\epsilon_n$ will appear small despite the solution being far from a periodic state due to their slow convergence. As a remedy, we propose a method in Section~\ref{sec_3D_periodicity} to estimate the asymptotic error $\epsilon_\infty$ as the simulation is being computed.

% Since the ratio $\alpha$ can be computed analytically, the asymptotic pressure $P_\infty$ can now be estimated from the pressure difference in two cardiac cycles
% %
% \begin{align}
% \bar{P}_\infty = \alpha \bar{P}_{n+1} + (1 - \alpha) \bar{P}_n
% \end{align}
% %
% This allows us to compute the asymptotic error from knowing only the solution of two cardiac cycles as
% %
% \begin{align}
% \epsilon_\infty = \frac{1}{1 + 1/\left[\alpha \left( \bar{P}_{n+1}/\bar{P}_n - 1\right)\right]}.
% \label{eq_err_asym_diff}
% \end{align}
% %
% Equation~\eqref{eq_err_asym_diff} can be used to determine whether a simulation has reached the desired tolerance $\epsilon_\infty$ from the solution of two consecutive cardiac cycles without explicit knowledge of the asymptotic pressure $\bar{P}_\infty$. 

\subsection{Checking the periodicity of 3D simulations \label{sec_3D_periodicity}}
% \subsection{Transition from analytical convergence analysis of single, independent 0d RCR BC to numerical convergence analysis of 3d model with multipled coupled 0d RCR BCs\label{sec_rcrc}}
% \subsection{Checking convergence of 3D simulation results \label{sec_check_3d_0d}}

The previous section introduced analytical and theoretical methods to check the periodicity of a single RCR model subjected to a prescribed inflow condition. This section extends this discussion to consider computational vascular models and present a method to check the periodicity of patient-specific, multi-outlet 3D models, where each outlet is coupled to a different RCR boundary condition. The steps for this method are summarized in the flowchart shown in Figure~\ref{fig_flowchart}. 

As mentioned in the previous section, $\bar{P}_\infty$ generally cannot be predicted analytically. However, a value of $\bar{P}_\infty$ is needed to check if the simulated flow rate and pressure at each outlet of the 3D model have converged to a periodic state. To remedy this issue, we estimate $\bar{P}_\infty$ for each outlet by simulating a separate 0D model of the attached RCR boundary condition. The inflow to each 0D RCR model is the simulated 3D flow rate, $Q_{3D}$, corresponding to that outlet. We then simulate these simple 0D RCR models for many cardiac cycles, to guarantee periodic convergence of the 0D RCR model, and use the final simulated pressure values, $P_{0D}$, as our estimates of $\bar{P}_\infty$. 

Furthermore, as previously discussed, $\epsilon_\infty$ represents the asymptotic error of the pressure solution. We therefore specify a desired value for $\epsilon_\infty$ and use this as our criteria for periodic convergence. Our 3D pressure solution, $P_{3D}$, is considered to be periodically converged, as per Equation~\eqref{eq_pressure_error_tolerance}, if it matches $P_{0D}$ within this error threshold. Mathematically, this condition is expressed as,
\begin{align}
\frac{P_{3D} - P_{0D}(Q_{3D})}{P_{0D}(Q_{3D})} \leq \epsilon_\infty.
\label{eq_3d_convergence}
\end{align}
Note that $Q_{3D}$ should be reasonably close to a periodic state before we can use it as the inflow to our RCR model. Typically, $Q_{3D}$ converges much faster, i.e., within one or two cardiac cycles, than $\bar{P}_{3D}$ does. This note will be further discussed in Section \ref{sec_res_conv}.

Lastly, each outlet of our 3D models has in general a unique time constant $\tau$. Although, in practice, parameter tuning yields time constants that are similar. It can be shown that the convergence of each outlet is determined by a single model time constant $\bar{\tau}$ that can be approximated by the mean of all individual time constants. As such, we can use this average time constant to estimate the number of cardiac cycles for which our 3D models must be simulated to achieve periodic convergence.

\subsection{Generating initial conditions \label{sec_ini}}
As previously discussed, 3D models must be simulated until the flow rates and pressures converge to a periodic state before they can be applied in scientific or clinical investigations, which currently requires running the simulation for several cardiac cycles. This limits the utility of computational 3D models and simulations in real-world clinical applications. To alleviate this bottleneck, we introduce a novel method to jumpstart the initialization of 3D simulations. %As we will demonstrate, this method significantly reduces the number of cardiac cycles that must be simulated before the 3D simulations converge to a periodic state, in contrast to the traditional methods of 3D initialization. % a method to reduce the computational expense of simulating 3D models. In particular, we present 
% Two sets of initial conditions commonly used in the vascular simulation community are 
In particular, we generate initial conditions to minimize the number of cardiac cycles required to reach periodic convergence within a chosen tolerance. This process is fully automated in SimVascular \cite{updegrove16} and requires no user input. Starting from a periodic 1D solution (Section~\ref{sec_ini_1d}) we create a map from the centerline, the 1D representation of the model, to the 3D volume mesh (Section~\ref{sec_ini_map}), and generate an artificial initial velocity vector field (Section~\ref{sec_ini_velocity}) for simulation initialization.

\subsubsection{Generating a periodic 1D solution \label{sec_ini_1d}}
We automatically generate a 1D flow model of the high-fidelity 3D flow model using the SimVascular 1D-plugin. We then run the model until it achieves periodic convergence and extract the solution of the last cardiac cycle.

\subsubsection{Mapping centerline to volume mesh \label{sec_ini_map}}
\begin{figure}[hbt!]
\centering
\includegraphics[width=0.15\linewidth]{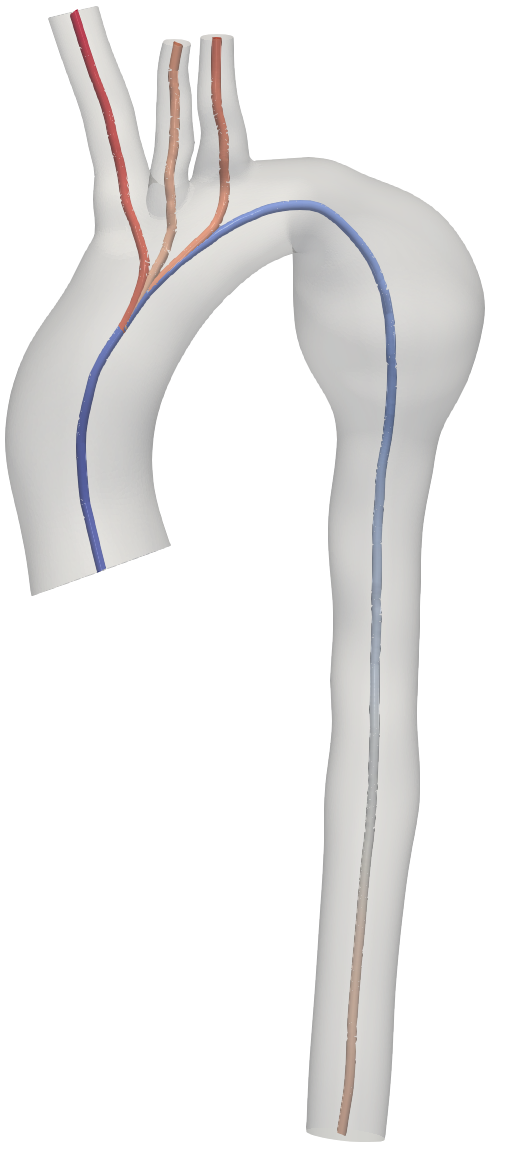}~
\includegraphics[width=0.15\linewidth]{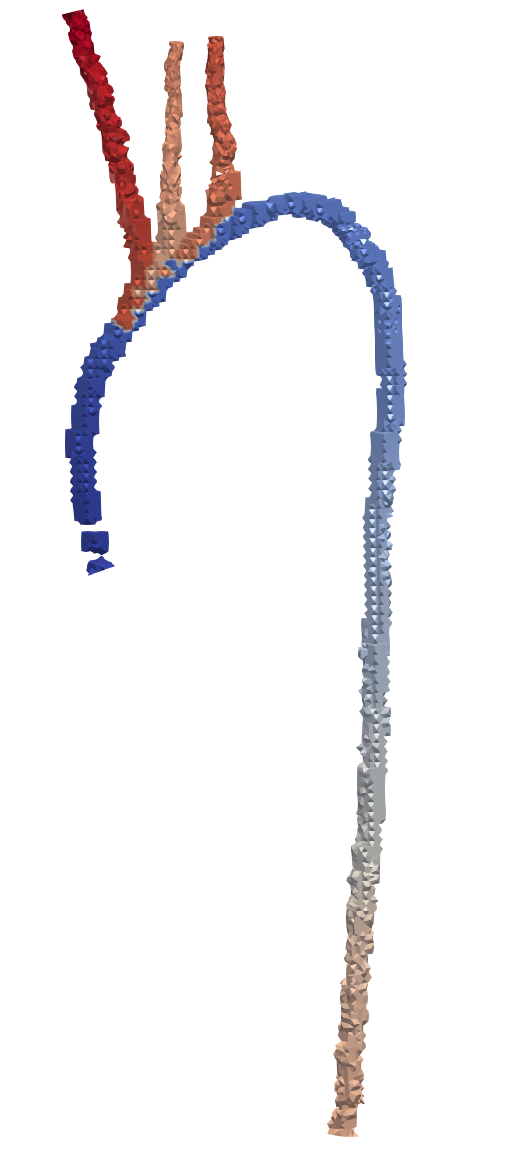}
\includegraphics[width=0.15\linewidth]{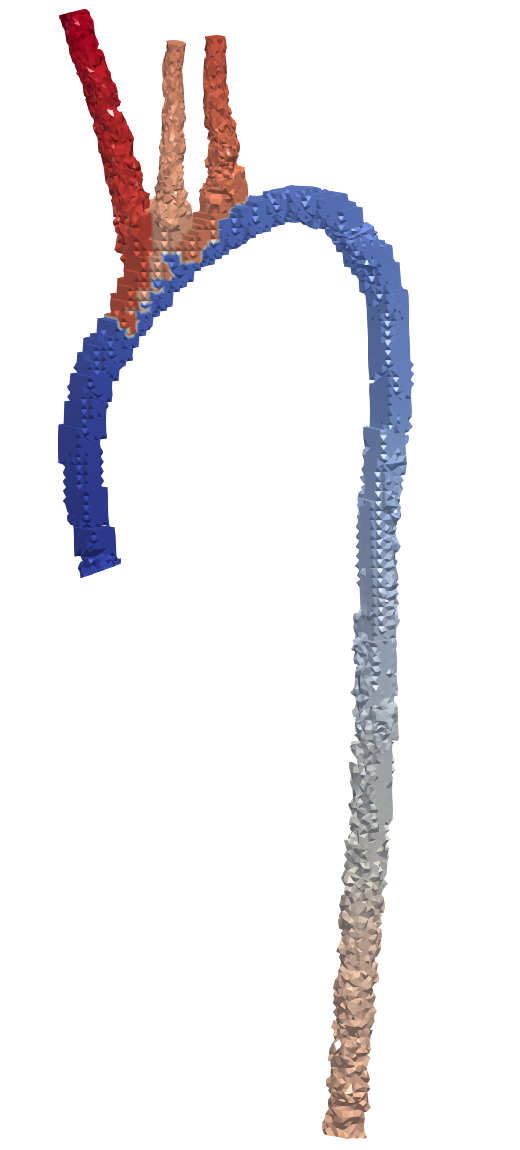}~
\includegraphics[width=0.15\linewidth]{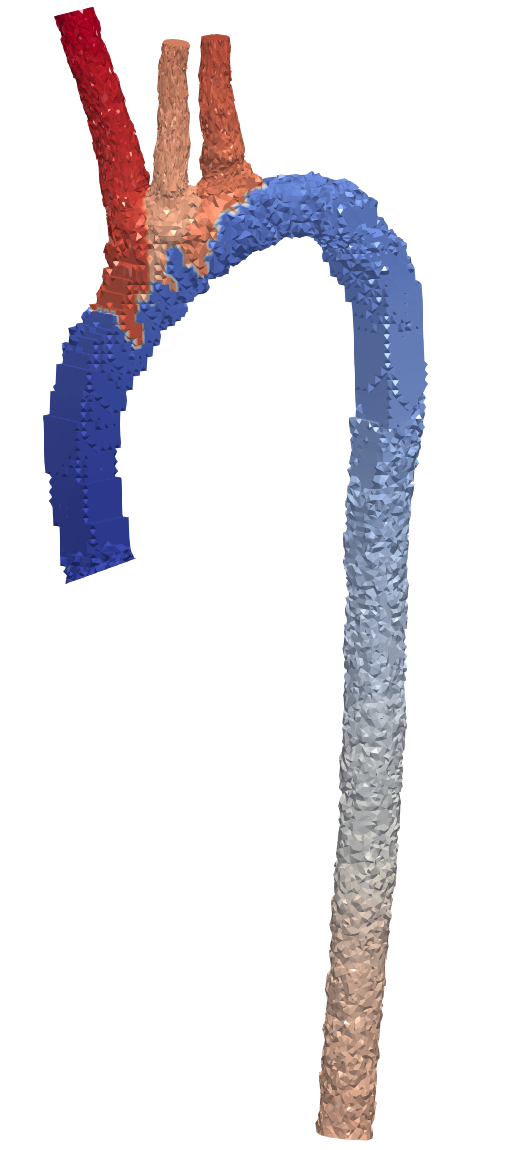}~
\includegraphics[width=0.15\linewidth]{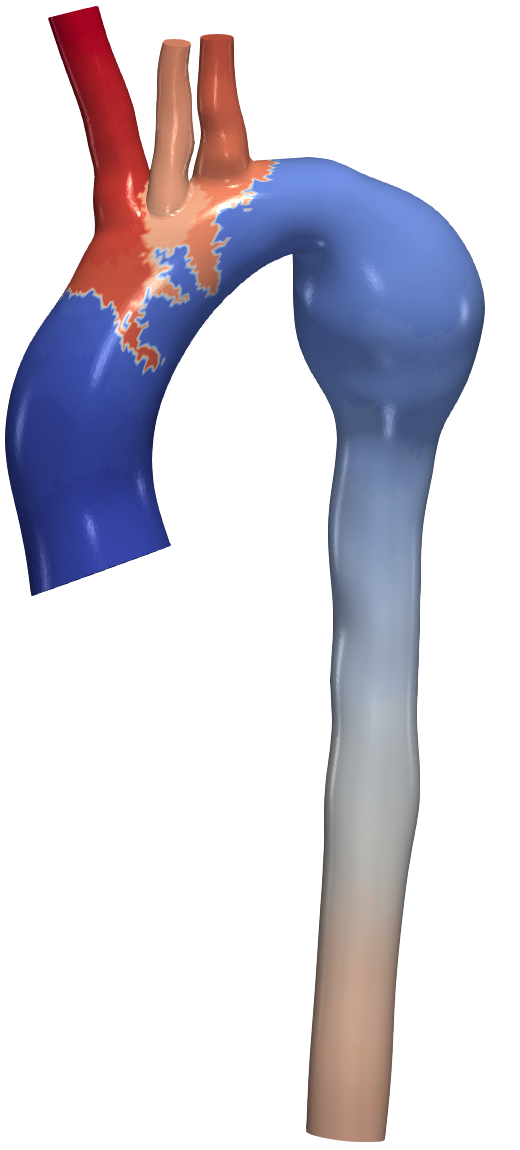}
\caption{Mapping centerline nodes to nodes in the 3D volume mesh. Colors correspond to the node order of the centerline. The centerline within 3D volume mesh is shown on the left. The figures from left to right show the current set of all points $P_\text{all}$ for iterations $i\in\{2,3,5,24\}$ until Algorithm~\ref{algo_map} converges.\label{fig_extrapolation}}
\end{figure}

To map the 1D solution to the 3D Finite Element mesh, we create a map $\vec{I}$ from nodes $P_\text{1D}$ on the 1D centerline to nodes $P_\text{3D}$ in the 3D volume mesh. This allows us to map quantities defined on the centerline to the volume mesh, such as 1D flow and pressure, cross-sectional area, and normal vectors. The iterative process is outlined in Algorithm~\ref{algo_map} and visualized in Figure~\ref{fig_extrapolation}. In Line~ \ref{algo_map_seed}, we first create a set of seed points $P_\text{seed}$ consisting of volume mesh nodes $P_\text{3D}$ that are closest to the centerline nodes $P_\text{1D}$ and store the corresponding indices in $\vec{I}$. To do this, we find the shortest Euclidean distances between centerline coordinates $\vec{x}^\text{1D}_p$ and 3D mesh coordinates $\vec{x}^\text{3D}_j$. We then employ a region growing algorithm (Line~\ref{algo_grow}) to grow the 1D-3D map outwards, starting from the centerline seed points $P_\text{seed}$. The algorithm in Line~\ref{algo_grow_step} selects nodes $P_\text{new}$ in the new layer from the previous layer $P_\text{old}$ using cell-connectivity. Finally, the map is expanded in Line~\ref{algo_ids} by assigning nodes in the new layer $P_\text{new}$ to the same 1D node as the closest 3D node in the previous layer $P_\text{old}$. This results in centerline nodes being roughly assigned to 3D mesh nodes within the same cross-section. For 3D meshes with $\mathcal{O}(10^6)$ nodes Algorithm~\ref{algo_map} only takes a few seconds to complete on a single CPU. We generate this map only once and use it repeatedly to map various centerline quantities to the volume mesh, see Figure~\ref{fig_velocity}.

\begin{algorithm}
\SetAlgoLined
\tcp{initialize empty map}
$\vec{I} \gets [0, \dots, 0]$\\
\tcp{initialize 3D seed points}
$P_\text{seed} \gets \{\}$\\
\tcp{loop all 1D points}
\For{$p \in P_\text{1D}$ \label{algo_map_seed}}{
\tcp{find closest 3D point}
$q = \underset{j \in P_\text{3D}}{\text{arg\,min}} ~ ||\vec{x}^\text{1D}_p - \vec{x}^\text{3D}_j||$\\
\If{$q \notin P_\text{seed}$}{
\tcp{add point to 3D seed points}
$P_\text{seed} \gets P_\text{seed} \cup \{q\}$\\
\tcp{assign map}
$\vec{I}_q = p$
}}
$C_\text{all} \gets \{\}$\\
$P_\text{all} \gets P_\text{seed}$\\
$P_\text{new} \gets P_\text{seed}$\\
$i \gets 0$\\
\tcp{in each iteration, grow seed points one layer outward}
\While{$|P_\text{new}| > 0 $ \label{algo_grow}}{
$P_\text{old} \gets P_\text{new}$\\
$P_\text{new} \gets \{\}$\\
\tcp{loop all 3D points in previous layer}
\For{$p \in P_\text{old}$ \label{algo_grow_step}}{
\tcp{loop all 3D cells connected to point}
\For{$c \in$ {PointCells}$(p)$}{
\tcp{skip 3D cells in previous layers}
\If{$c \notin C_\text{all}$}{
$C_\text{all} \gets C_\text{all} \cup \{c\}$\\
\tcp{loop all 3D points connected to cell}
\For{$q \in \text{CellPoints}(c)$}{
\tcp{skip 3D points in previous layers}
\If{$q \notin P_\text{all}$}{
\tcp{add 3D points to new layer}
$P_\text{new} \gets P_\text{new} \cup \{q\}$\\
$P_\text{all} \gets P_\text{all} \cup \{q\}$\\
}}}}}
\tcp{loop all 3D points in new layer}
\For{$p \in P_\text{new}$ \label{algo_ids}}{
\tcp{assign map according to map of closest 3D point in previous layer}
$\vec{I}_p = \vec{I}_q \text{~with~} q =  \underset{j \in P_\text{old}}{\text{arg\,min}} ~ ||\vec{x}^\text{3D}_p - \vec{x}^\text{3D}_j||$
}
$i \gets i+1$\\
}
\caption{Mapping centerline nodes to nodes in the 3D volume mesh. \label{algo_map}}
\end{algorithm}

\subsubsection{Extrapolating pressure and velocity \label{sec_ini_velocity}}
\begin{figure}[hbt!]
\centering
\includegraphics[width=0.15\linewidth]{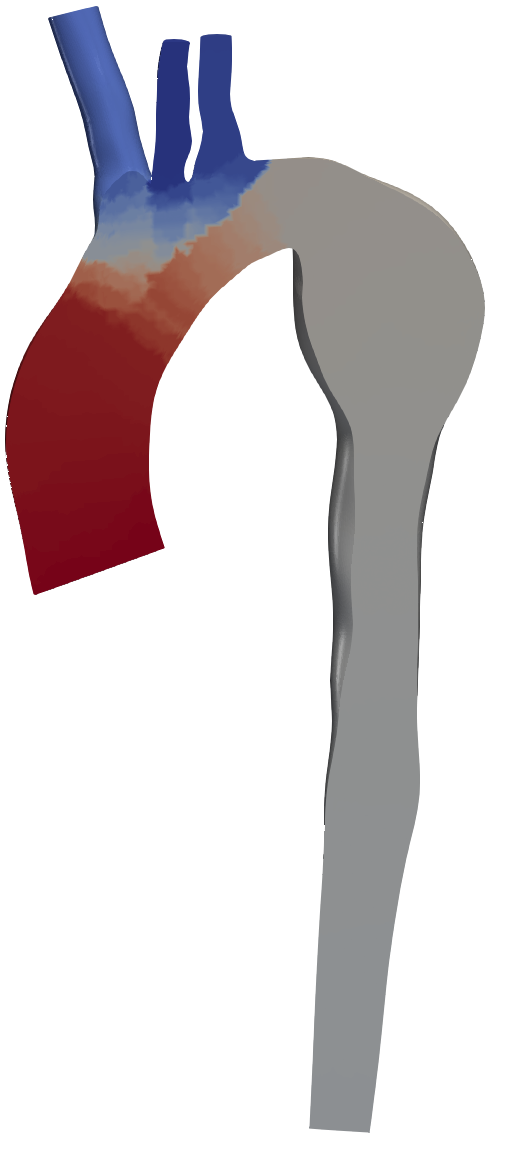}
\includegraphics[width=0.15\linewidth]{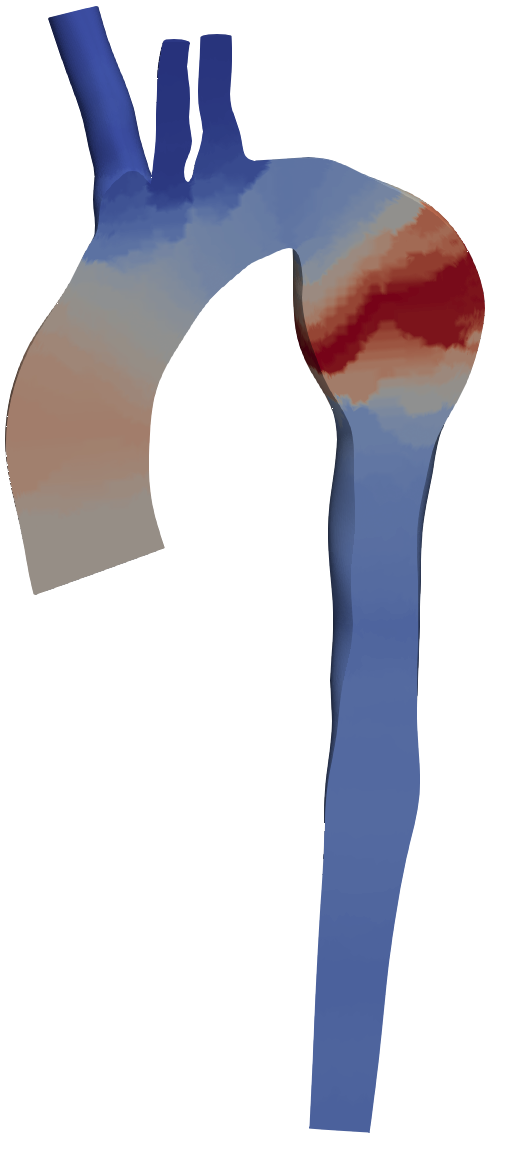}~
\includegraphics[width=0.15\linewidth]{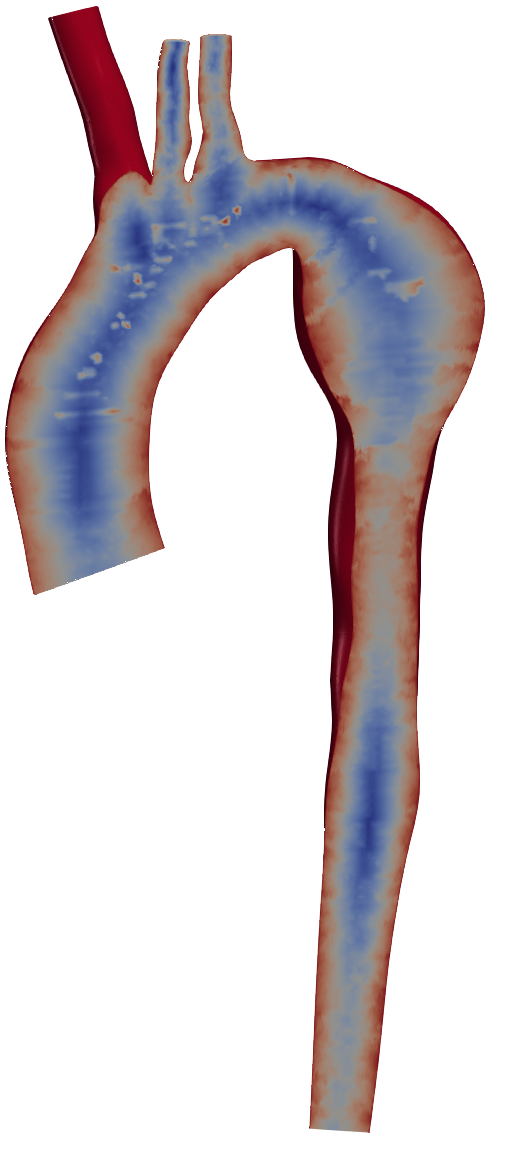}~
\includegraphics[width=0.15\linewidth]{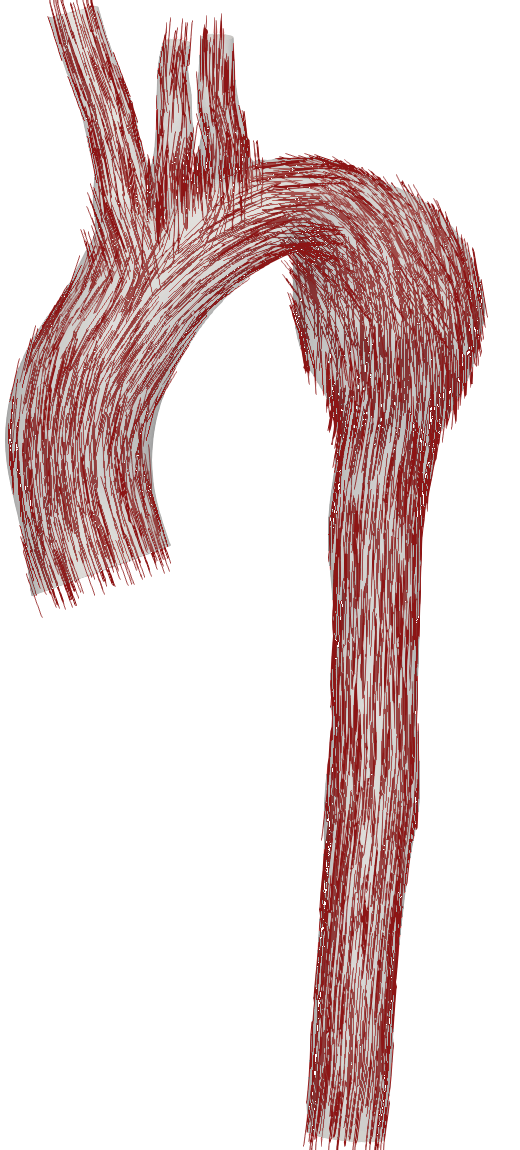}
\caption{Ingredients to generate the velocity field from a 1D solution. From left to right: Flow mapped from 1D, cross-sectional area, radial coordinate, and normal vectors\label{fig_velocity}}
\end{figure}

We directly map the pressure from the centerline to the volume mesh using the map generated in Section~\ref{sec_ini_map}, resulting in a pressure that's approximately constant over the cross-section of the vessels. However, as the 1D solution only provides a scalar flow along the centerline, we must generate a velocity vector field from scratch. The ingredients for the velocity vector field are visualized in Figure~\ref{fig_velocity}. From the mapped 1D flow rate and the cross-sectional area  we calculate the velocity magnitude. Assuming a Poiseuille flow, we apply a parabolic flow profile to the velocity magnitude. With the help of a normalized radial coordinate, we prescribe the flow profile to be maximal on the centerline and zero on the boundary. Finally, we multiply the scalar velocity magnitude with the centerline tangent vector to generate a vector field. We aim to preserve the amount of flow through a cross-section of the vessel as obtained from the 1D solution while approximating a somewhat physiological velocity field.  This velocity vector field can then be used for simulation initialization.

\section{Results \label{sec_res}}
Throughout this section, we consider a simulation periodically converged if the asymptotic error is $\epsilon_\infty\leq1\,\%$, as defined in Equation~\eqref{eq_err_asym}. We utilize in this work a subset of 52 models from the Vascular Model Repository (\url{vascularmodel.org}) \cite{wilson13}. Six out of these models are shown in Table~\ref{tab_models}.

\begin{table}[hbt!]
\centering
\begin{tabularx}{\linewidth}{|s|Y|Y|Y|Y|Y|Y|}
\hline
& \includegraphics[height=3cm]{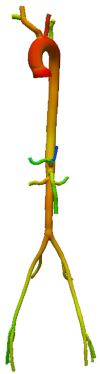} 
& \includegraphics[height=1.5cm]{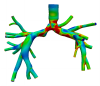} 
& \includegraphics[height=3cm]{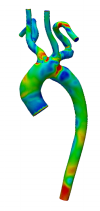}
& \includegraphics[height=3cm]{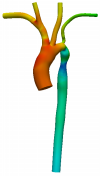}
& \includegraphics[height=3cm]{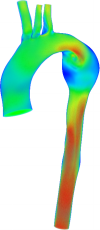}
& \includegraphics[height=3cm]{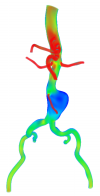}\\
\hline
\textbf{ID} & 0003\_0001 & 0097\_0001 & 0107\_0001 & 0111\_0001 & 0130\_0000 & 0156\_0001\\ 
\hline
\textbf{Type} & aorta-femoral & pulmonary & aorta & aorta & aorta & aorta-femoral\\
\hline
\textbf{State} & normal & Glenn & end-to-end anastomosis & coarctation & aneurysm & aneurysm\\
\hline
\end{tabularx}
\caption{Subset of models from the Cardiovascular Model Repository.\label{tab_models}}
\end{table}

%\begin{align}
%\centering
%\epsilon_P = \frac{1}{n_t} \sum_{j=1}^{n_t} \left| \frac{P_{j}^\infty - P_{j}}{P_{j}^\infty} \right|,  \quad
%\epsilon_{Q} = \frac{1}{n_t} \sum_{j=1}^{n_t} \left| \frac{Q_{j}^\infty - Q_{j}}{\max_k \left( \left| Q_{k}^\infty \right| \right) } \right|
%\label{eq_error_def}
%\end{align}

\subsection{Convergence}
In this section, we detail the convergence behavior of cardiovascular fluid dynamics simulations with three-element Windkessel, or RCR, boundary conditions. The models used in this section have different numbers of outlets, ranging from 4 in the aortic models to 33 in the pulmonary artery models. For each model, we automatically generate a reduced-order 0D model as outlined in Section~\ref{sec_mod_0d}. Using a computationally inexpensive 0D model allows us to run a large number of simulations for many cardiac cycles and analyze their convergence behavior in detail. These 0D results are directly applicable to 1D and 3D simulation models and form the foundation for generating initial conditions in Section~\ref{sec_res_ini}.

\subsubsection{Time constants}
\begin{figure}[hbt!]
\centering
\includegraphics[width=\linewidth]{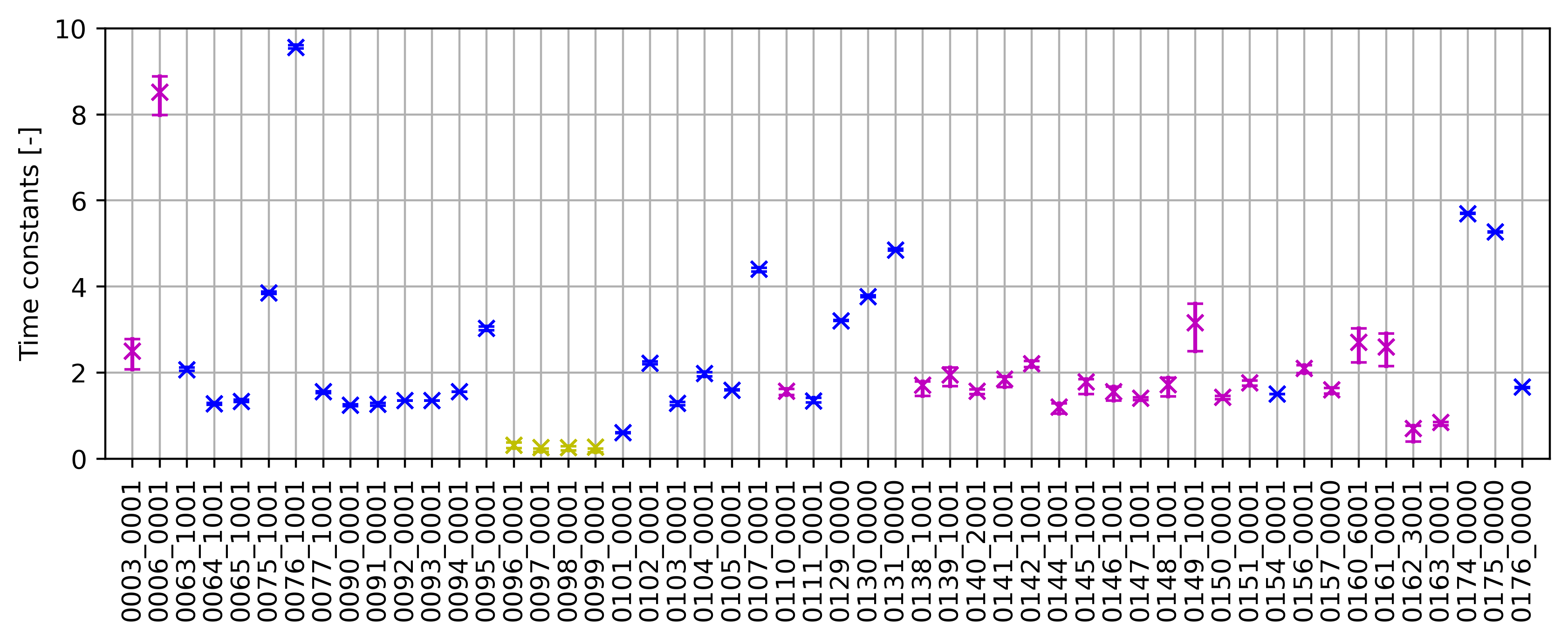}
\caption{Time constants for models from the Vascular Model Repository. Bars indicate the range of time constants $\tau/T$ for all outlets, crosses indicate the model time constant $\bar{\tau}/T$. The colors indicate the model category: aorta (blue), aorta-femoral (purple), and pulmonary (yellow).\label{fig_time_constants}}
\end{figure}

The time constant $\tau$ is the metric of a boundary condition that determines the rate of periodic convergence. It is thus essential for all numerical experiments in this work. Throughout the remainder of this work, we normalize $\tau$ by the length of the cardiac cycle $T$, with all values reported in Figure~\ref{fig_time_constants}. All normalized time constants fall within the range $[0,10]$, with time constants in pulmonary models being the lowest. Furthermore, the time constants of different outlets in one model all fall within a narrow range. In addition to the outlet time constants, we also show the model time constants $\bar{\tau}/T$ as crosses. We extract the model time constants from the slope of the exponential curves of the asymptotic error $\epsilon_\infty$ (see Equation~\eqref{eq_err_asym}) when running the models for several cardiac cycles. Each model has one unique model time constant $\bar{\tau}/T$ that is approximately the mean value of the outlet time constants.

\subsubsection{Initial conditions}
\begin{figure}[hbt!]
\centering
\begin{subfigure}{.9\linewidth}
\includegraphics[width=\linewidth]{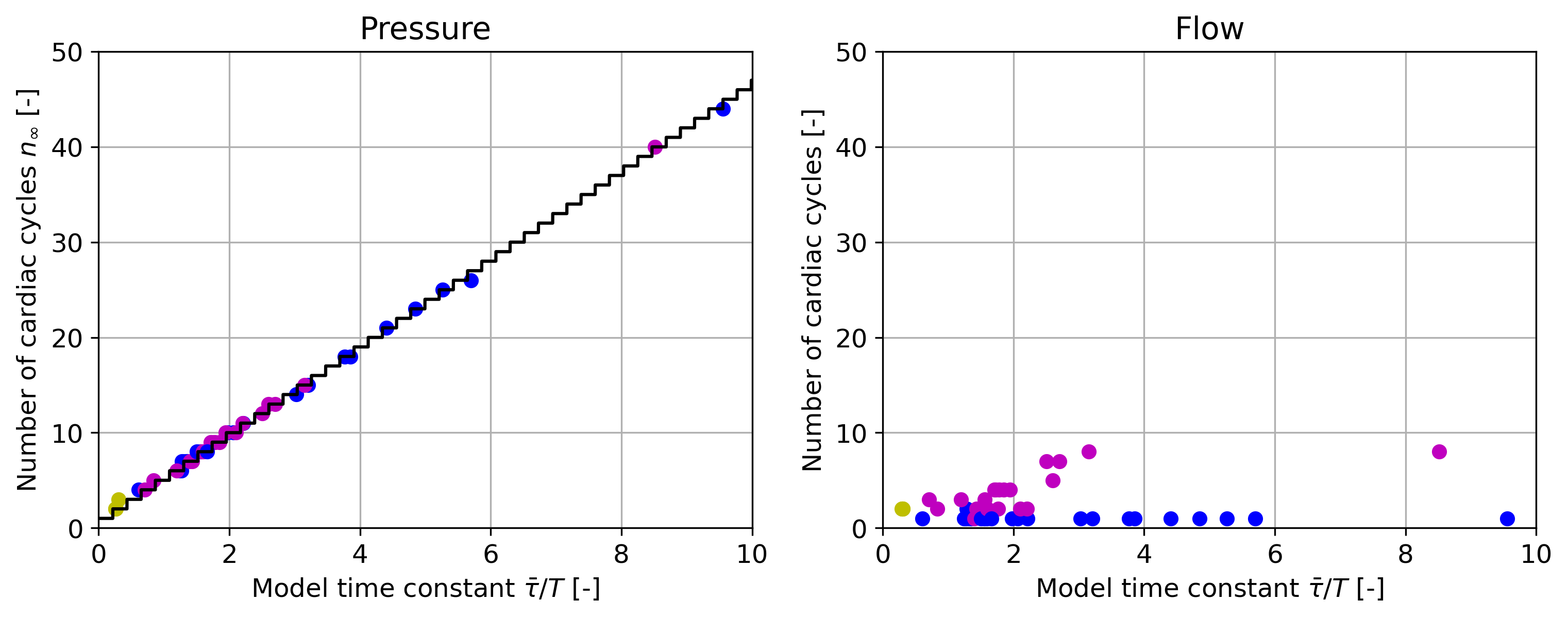}
\caption{Starting from zero initial conditions\label{fig_zero_initial}}
\end{subfigure}
\begin{subfigure}{.9\linewidth}
\includegraphics[width=\linewidth]{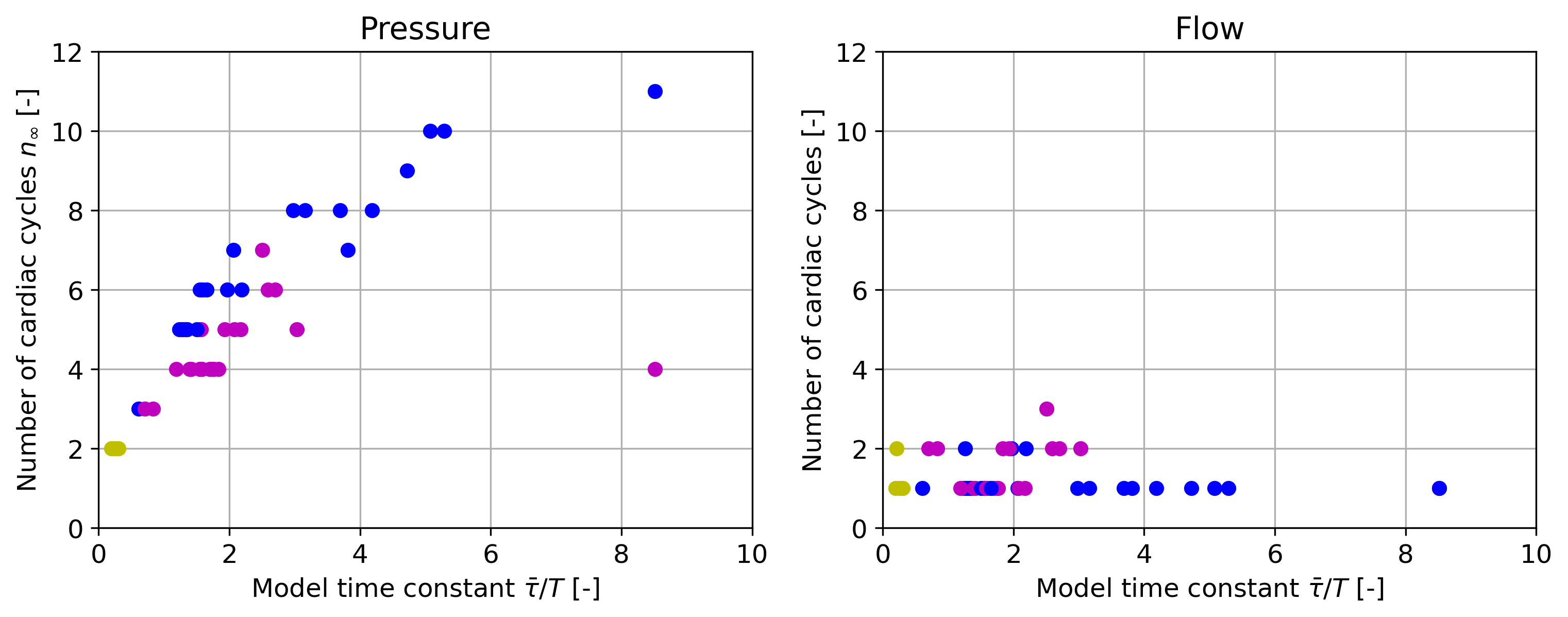}
\caption{Starting from steady state initial conditions\label{fig_steady_initial}}
\end{subfigure}
\caption{Number of cardiac cycles $n_\infty$ required to reach an asymptotic error $e_\infty=1\,\%$ for 0D models from the Vascular Model Repository. The colors indicate the model category: aorta (blue), aorta-femoral (purple), and pulmonary (yellow). The black line (left) indicates the number of cardiac cycles predicted by Equation~\eqref{eq_err_asym_zero}.}
\end{figure}

We estimated the number of cardiac cycles $n_\infty$ required to reach a periodic pressure in Equation~\eqref{eq_err_asym}. In the special case of zero initial conditions, this relationship simplified to the inequality in Equation~\eqref{eq_err_asym_zero}, scaling linearly with the time constant $\tau/T$ and the logarithm of the asymptotic error $\epsilon_\infty$. This relationship holds regardless of model fidelity, i.e. for 0D, 1D, and 3D. In Figure~\ref{fig_zero_initial}, we record the number of cardiac cycles required to reach an asymptotic error of $\epsilon_\infty\leq1\,\%$ when starting a simulation from zero initial conditions, and report the number of cardiac cycles required for periodic pressure (left) and flow (right) over the model time constant $\bar{\tau}/T$. For pressure (left), the results confirm the linear relationship between the single model time constant $\bar{\tau}/T$ and the number of cardiac cycles $n_\infty$ from  Equation~\eqref{eq_err_asym_zero} (black line). For a range of model time constants $\bar{\tau}/T \ \in [0.3, 9.6]$ we find the range of cardiac cycles $n_\infty \in [2, 44]$. The number of cardiac cycles to reach a periodic flow solution (right) is not correlated to the model time constant and is, in general, much lower than the number of cycles $n_\infty$ to reach a periodic pressure. Flow rate commonly converges within one cardiac cycle, with a maximum of eight cardiac cycles in our cohort of models. 

The same study is repeated in Figure~\ref{fig_steady_initial} for starting from steady state initial conditions. Compared to zero initial conditions, the number of cardiac cycles required to each an asymptotic pressure with an error of $\epsilon_\infty\leq1\,\%$ are much lower: $n_\infty \in [2, 12]$. Flow is converged within a maximum of 5 cardiac cycles. When starting from steady state initial conditions, the number of cardiac cycles $n_\infty$ cannot be given analytically, it depends on the prescribed inflow profile. However, the number of cardiac cycles commonly still increases with the model time constant.

\subsubsection{Comparison of error metrics \label{sec_res_conv}}
\begin{figure}[hbt!]
\centering
\includegraphics[width=\linewidth]{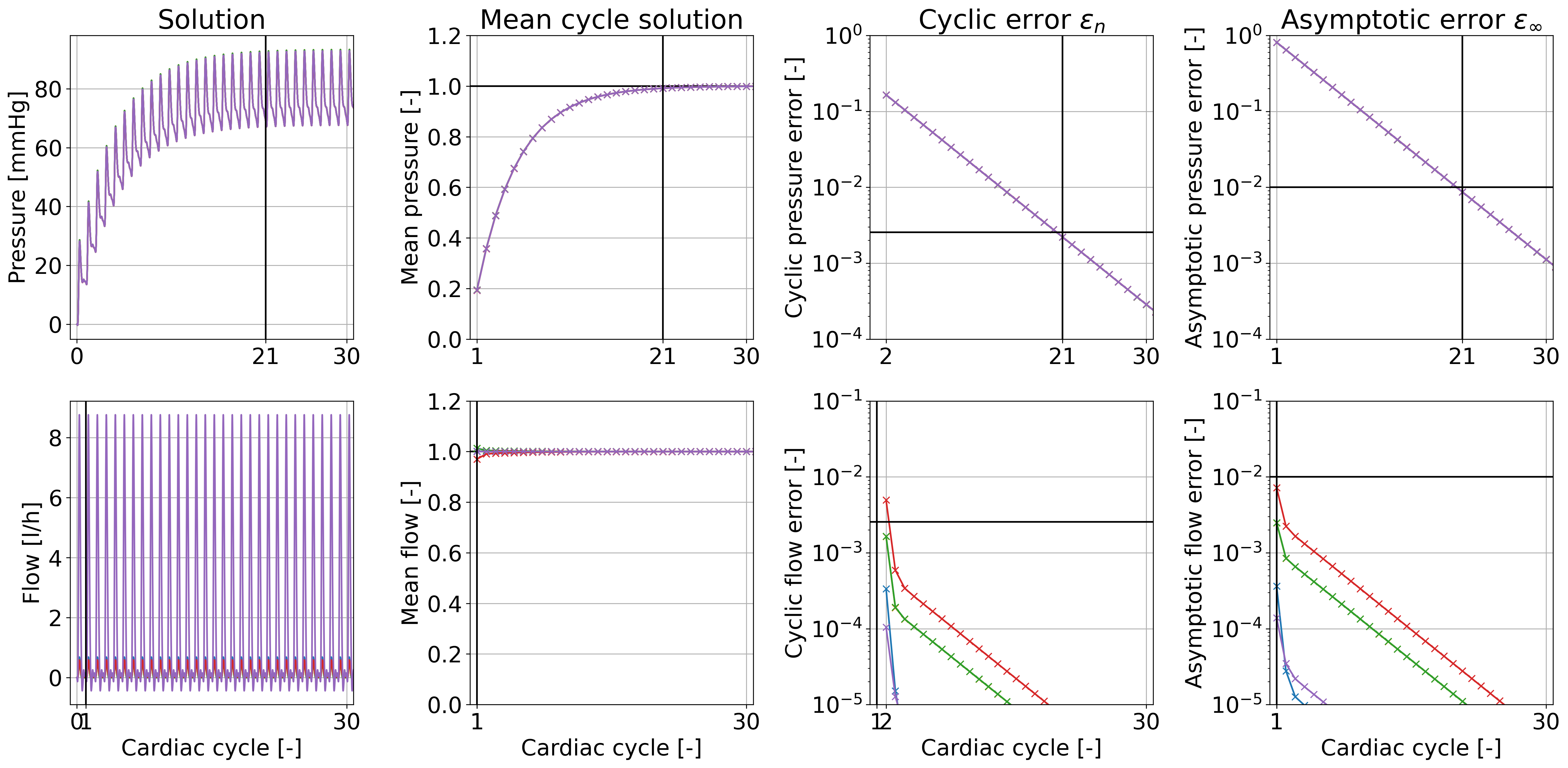}
\caption{Convergence of 0D pressure (top) and flow (bottom) solution in geometry 0107\_0001 (normal aorta) for $n=30$ cardiac cycles. From left to right: Solution, mean solution over one cardiac cycle (normalized by asymptotic solution), cyclic error $\epsilon_n$, and asymptotic error $\epsilon_\infty$. Each color refers to one of the outlets.\label{fig_res_conv_0d}}
\end{figure}

We visualize the convergence of pressure and flow for $n=30$ cardiac cycles in Figure~\ref{fig_res_conv_0d} for model 0107\_0001 (normal aorta). The pressure curve (top left) builds up slowly in each cardiac cycle, starting from zero, whereas the flow curve (bottom) is close to periodic starting from the first cycle. Taking the mean value over each cardiac cycle, both solutions exponentially approach their periodic state, as shown in Equation~\eqref{eq_step_puls}. The logarithmic plots in the two rightmost columns show the exponential decay of the cyclic error $\epsilon_n$ and the asymptotic error $\epsilon_\infty$, as defined in Equations~\eqref{eq_err_asym} and \eqref{eq_err_cyc}, respectively. Note that the flow (bottom) exhibits smaller errors than the pressure (top) and converges faster during the first few cardiac cycles. After that, both flow and pressure at all outlets converge with the model time constant $\bar{\tau}/T \approx 4.4$. This model time constant yields a factor between cyclic and asymptotic error of $\alpha = \epsilon_\infty/\epsilon_n\approx 3.9$. The threshold is indicated by horizontal lines in the error plots. Here, the solution is converged after $n_\infty=21$ cardiac cycles, reaching errors of $\epsilon_\infty = 1\,\%$ and $\epsilon_n = \epsilon_\infty/\alpha \approx 0.26\,\%$. This example demonstrates that for simulations with $\bar{\tau}/T > 1/\ln 2 \approx 1.44$ the cyclic error $\epsilon_n$ underestimates the asymptotic error $\epsilon_\infty$.

\subsubsection{Prediction of the periodic state}
\begin{figure}[hbt!]
\centering
\includegraphics[width=\linewidth]{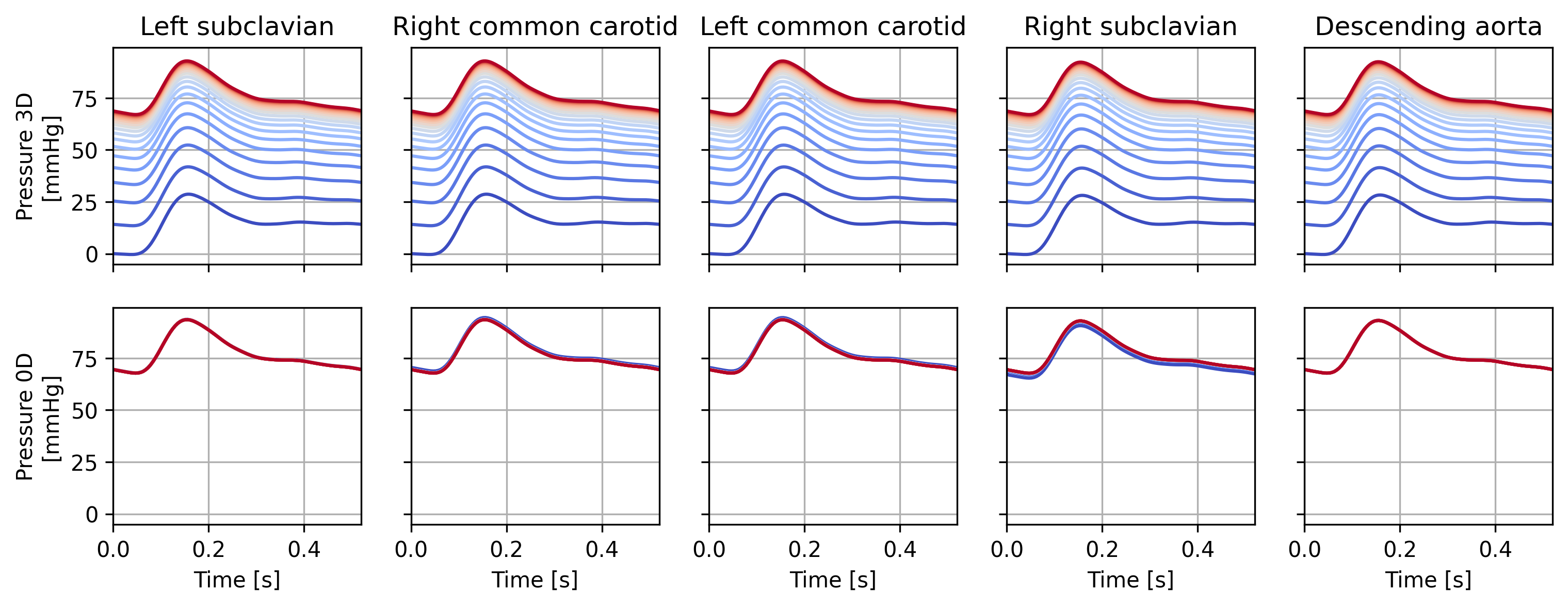}
\caption{Pressure for model 0107\_0001 for $n_\infty=21$ cardiac cycles (blue to red) starting from zero initial conditions until reaching periodic convergence. Pressure at the 3D outlets (top) and peridioc state predicted from 0D (bottom). \label{fig_res_conv_online}}
\end{figure}

As explained in Section~\ref{sec_rcrc}, the periodic error $\epsilon_\infty$, comparing the current cardiac cycle to a perfectly periodic cycle, can in general not be computed analytically. It requires the periodic solution of the model which is not known \emph{a priori}. We thus outlined a method in Section~\ref{sec_3D_periodicity} to use the 0D lumped-parameter boundary condition to estimate the periodic cycle numerically. Figure~\ref{fig_res_conv_online} shows the pressure at all outlets of model 0107\_0001 from cycle one to cycle $n_\infty=21$ (from blue to red), starting from zero initial conditions as in Figure~\ref{fig_res_conv_0d}. The top row shows the pressure at the outlets of the 3D model in each cardiac cycle. The bottom row shows the prediction of the periodic state using the flow at the outlets of the 3D model in each cardiac cycle. Since flow converges much faster than pressure in this model, see Figure~\ref{fig_res_conv_0d}, the cycle-to-cycle variation is minimal. The periodic state can be accurately predicted even from early cardiac cycles, where the actual pressure of the model has not yet converged.

\subsection{Initialization}
To demonstrate the performance of our 3D initialization method, we compare results for the 6 models shown in Table~\ref{tab_models}. The chosen models contain a wide range of anatomies and physiological conditions and are specified by an eight-digit ID. They include aorta and femoral arteries (0003\_0001), pulmonary arteries after a Glenn procedure (0097\_0001), aortic coarctation post end-to-end anastomosis (0107\_0001), untreated aortic coarctation (0111\_0001), aortic aneurysm in a patient with Marfan Syndrome (0130\_0000), and abdominal aortic aneurysm (0156\_0001).

\subsubsection{Generating initial conditions \label{sec_res_ini_cond}}
In this section, we demonstrate the performance of our pipeline to generate 3D initial conditions that greatly reduce the number of cardiac cycles, as proposed in Section~\ref{sec_ini}. As a ground truth, i.e. the ideal initial condition, we use an asymptotic 3D solution extracted  after reaching periodic convergence. Using this initial condition, the 3D simulation would reach periodic converge within one cardiac cycle. We compare this solution to initial conditions we generated from a periodic 1D solution, using the mapping technique from Section~\ref{sec_ini_map}. By using initial conditions mapped from 1D to 3D, we introduce two kinds of errors. First, the 1D solution is computed on a highly simplified geometrical representation of the 3D geometry assuming a constant flow profile, see Section~\ref{sec_mod_1d}. Second, the mapping process tries to but cannot guarantee local preservation of the 1D solution characteristics, such as incompressibility.

\begin{figure}[hbt!]
\centering
\includegraphics[width=\linewidth]{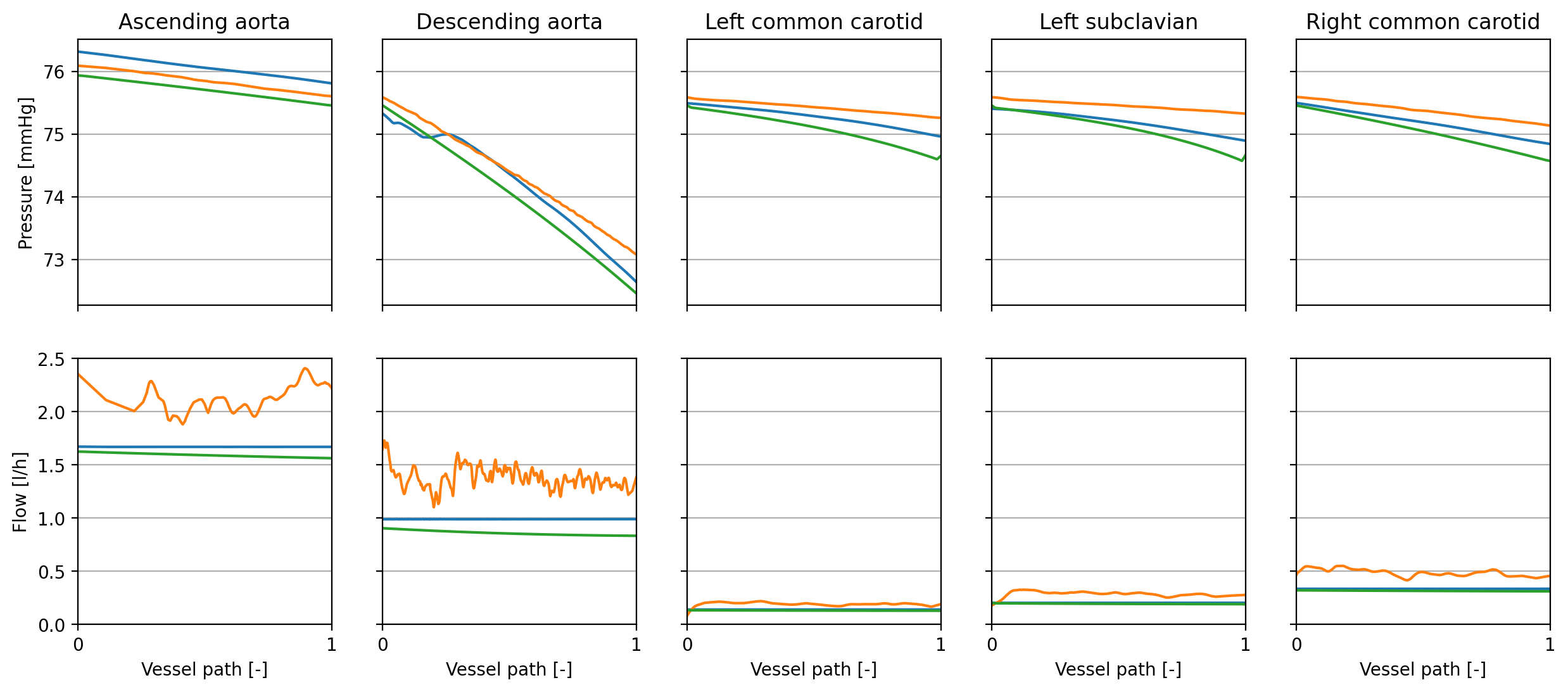}~
\caption{Initial conditions in aortic aneurysm model 0130\_0000 for different models: 3D ground truth (blue), 1D solution (green), 3D mapped from 1D (orange).\label{fig_ini_int}}
\end{figure}

Figure~\ref{fig_ini_int} shows these errors by comparing different initial conditions: 3D ground truth (blue), 1D solution (green), and 3D mapped from 1D (orange). We integrate the 3D ground truth and the mapped 3D initial conditions over the cross-section of the 3D geometry continuously along the centerline. This allows us to plot pressure (top) and flow (bottom) continuously over the vessel path. For a perfect 1D approximation of the 3D simulation, the blue and green lines would overlap. For a perfect mapping from 1D to 3D, the blue and orange lines would overlap. In general, pressure is approximated well by the 1D model and mapped well to the 3D domain. The differences between the models ($<1$\,mmHg) are small compared to the overall pressure level ($\sim 75$\,mmHg). Flow is approximated well by the 1D model whereas the the mapping results in oscillations and overestimates the actual flow. Since all three models should represent a model with rigid walls and incompressible flow, flow should be constant along the vessel path. However, the 3D mapped from 1D solution still roughly represents the correct flow splits to the different vessel branches.

\begin{figure}[hbt!]
\centering
\begin{subfigure}{0.3\linewidth}
\includegraphics[width=.5\linewidth]{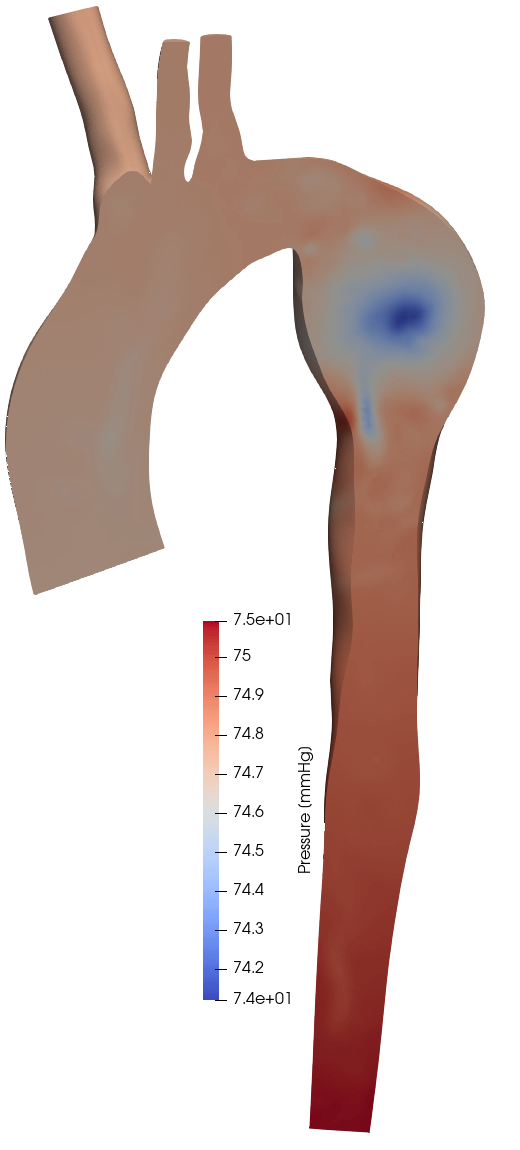}~
\includegraphics[width=.5\linewidth]{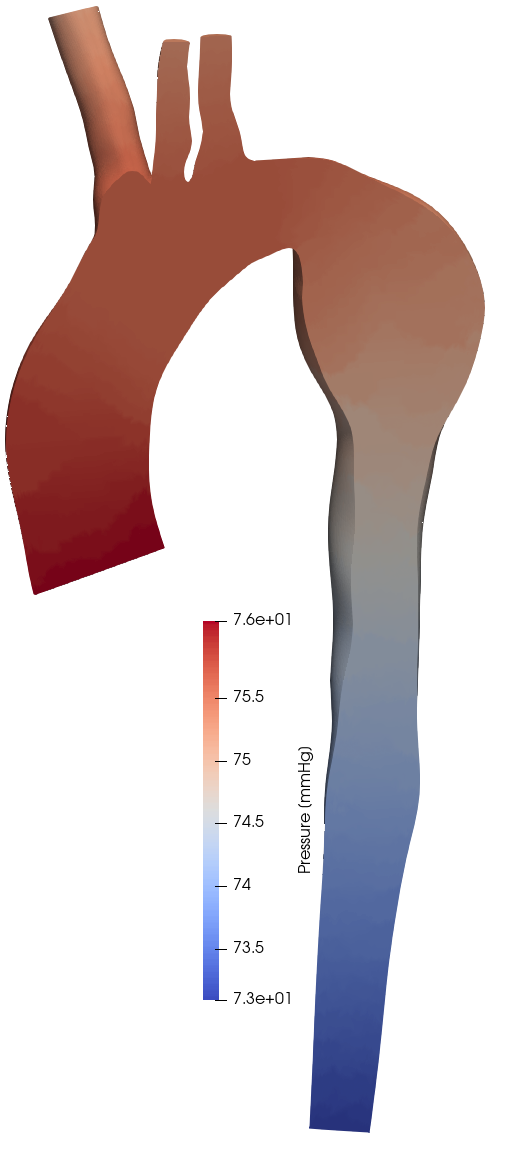}
\caption{Pressure: ground truth (left), mapped from 1D (right) \label{fig_ini_1d_3d_pressure}}
\end{subfigure}\hspace{2cm}
\begin{subfigure}{0.3\linewidth}
\includegraphics[width=.5\linewidth]{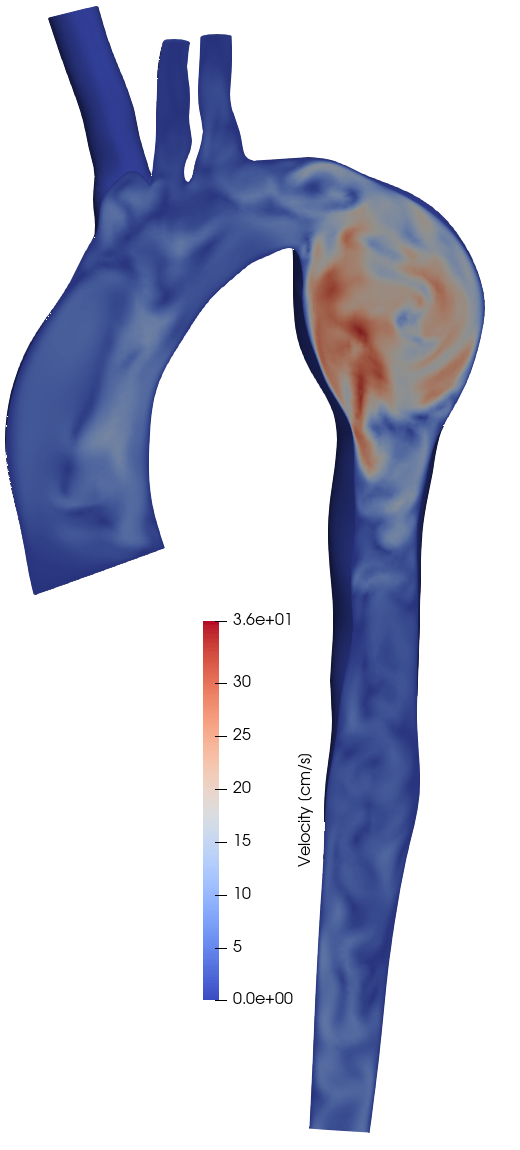}~
\includegraphics[width=.5\linewidth]{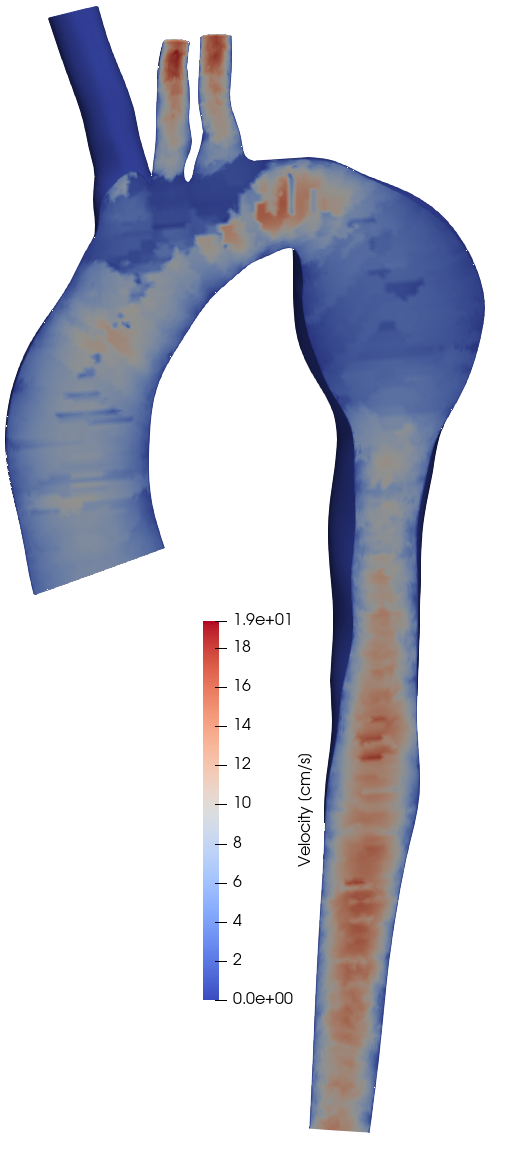}
\caption{Velocity (mag.): ground truth (left), mapped from 1D (right) \label{fig_ini_1d_3d_flow}}
\end{subfigure}
\caption{Initial conditions mapped from 1D compared to the 3D asymptotic solution in aortic aneurysm model 0130\_0000.\label{fig_ini}}
\end{figure}

The differences between 3D ground truth and 3D mapped from 1D initial conditions are visualized in Figure~\ref{fig_ini_1d_3d_pressure} and are evident, especially within the aneurysm. The mapping from 1D to 3D results in a pressure that is approximately constant over the cross-sections of the vasculature. However, the overall variation, 74-75~mmHg in the ground truth and 73-76~mmHg in the extrapolated solution, is negligibly small compared to the overall pressure level. Similarly, the velocity field in Figure~\ref{fig_ini_1d_3d_flow} is visibly different for the ground truth (left) and mapped solution (right).

\subsubsection{Initial conditions \label{sec_res_ini}}
Finally, we demonstrate that 3D initial conditions mapped from periodic 1D solutions can greatly reduce the 3D simulation time. We compare three kinds of initial conditions: Zero, Steady, and 1D. For zero initial conditions, we prescribe uniformly $P_0(\vec{x}) = \vec{0}$ and $\vec{v}_0 = 10^{-4} \cdot [1,1,1]$. To generate steady initial conditions, we first run a steady state simulation with constant mean inflow $\bar{Q}$. We compared starting the pulsatile flow simulation at different time steps for model 0003\_0001. Common choices are either the first time step $t=0$ or the time step where the inflow closely matches the mean flow \cite{vignonclementel10}. We found that the number of cardiac cycles required to reach periodic convergence was not affected by the choice of starting time step, so for simplicity we chose $t=0$ for all comparisons in this section. For 1D initial conditions, we use the framework proposed in Section~\ref{sec_ini}. The 3D simulations were run using an incompressible Newtonian fluid, a rigid wall, and a no-slip condition at the wall. All 1D simulations and the mapping from 1D to 3D were run a single CPU on a workstation computer.

\begin{figure}[hbt!]
\centering
\includegraphics[width=\linewidth]{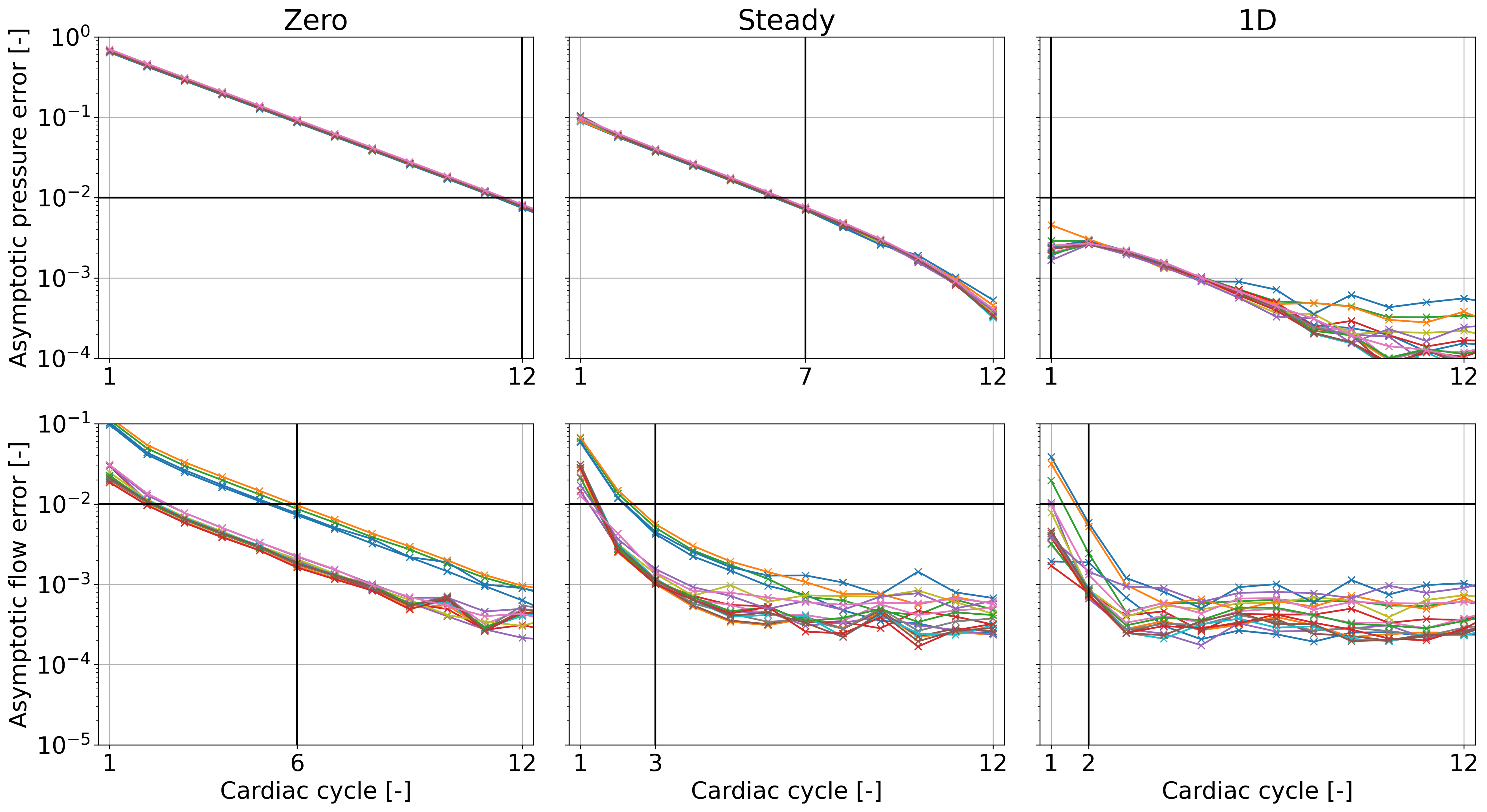}
\caption{Convergence of the 3D pressure in geometry 0003\_0001 (aorta-femoral) for different initial conditions: zero (left), steady (middle), 1D (right). It shows pressure (top) and flow (bottom) over multiple cardiac cycles. Each color refers to one of the outlets. The horizontal black line indicates the asymptotic error of $\epsilon_\infty\leq1\,\%$. The vertical black line indicates after how many cardiac cycles the simulation results have reached that asymptotic error.\label{fig_convergence}}
\end{figure}

Figure~\ref{fig_convergence} shows the convergence of the asymptotic error $\epsilon_\infty$ for pressure (top) and flow (bottom) at all outlets (colors) for all three initial conditions: zero (left), steady (middle), 1D (right). The threshold $\epsilon_\infty \leq 1\,\%$ is indicated by a horizontal line, the number of cardiac cycles $n_\infty$ required to reach that threshold is indicated in each simulation by a vertical line. All simulations are shown for a total of $n=12$ cardiac cycles. Note that the scale of the flow error is one magnitude lower than the scale of the pressure error. From left to right, the flow and pressure solutions start with a successively lower error in the first cardiac cycle. In the case of 1D initial conditions (right), the pressure errors of all outlets already fulfil the convergence criterion after only one cardiac cycle. Both pressure and flow errors stagnate below $10^{-3}$ as other numerical errors in the simulations outweigh the asymptotic error. As previously observed, pressure converges faster than flow when using zero or steady initial conditions. However, pressure and flow are converged in one and two cardiac cycles, respectively, when using 1D initial conditions. Due to the drastic reduction in the number of cardiac cycles required for the pressure solution, flow is now slightly slower to converge.

Convergence results for all selected 3D geometries and are reported in Table~\ref{tab_models_cycles}. While initializing with a steady state solution considerably reduces the number of cardiac cycles to reach convergence, most of the models still require at least five cardiac cycles for convergence. In all models, the simulations initialized with the 1D solution converge in one or two cardiac cycles. This leads to speedup factors between one and nine compared to the steady state initialization, saving between 0 and 53\,h of computation time on 48 CPU cores each per 3D model. Only in model 0097\_0001 which has a very low model time constant and the simulation even with zero initial conditions converges in two iterations does the 1D initialization not yield a speedup.

\begin{table}[hbt!]
\centering
\begin{tabularx}{\linewidth}{|s|Y|Y|Y|Y|Y|Y|}
\hline
\textbf{ID} & 0003\_0001 & 0097\_0001 & 0107\_0001 & 0111\_0001 & 0130\_0000 & 0156\_0001\\ 
\hline
\hline
$t_\text{3D}$ & 2.2\,h & 1.7\,h & 1.9\,h & 6.3\,h & 6.6\,h & 6.4\,h \\
\hline
\textbf{$\bar{\tau}/T$} & 2.5 & 0.2 & 4.4 & 1.3 & 3.8 & 2.1 \\
\hline
\hline
\textbf{Zero}   & \textbf{12} / 6 &  \textbf{2} / 2 & \textbf{20} / 1 & \textbf{7} / 1 & \textbf{16} / 1 & \textbf{10} / 3 \\
\hline
\textbf{Steady} & \,~\textbf{7} / 3  & \textbf{2} / 2 & \,~\textbf{6} / 1 & \textbf{5} / 2 & \,~\textbf{9} / 3 & \,~\textbf{5} / 2 \\
\hline
\textbf{1D}     & \,~1 / \textbf{2}  & 1 / \textbf{2} & \,~\textbf{1} / 1 & \textbf{1} / 1 & \,~\textbf{1} / 1 & \,~1 / \textbf{2} \\
\hline
\hline
$t_\text{saved}$  & 11\,h & 0 & 9.5\,h & 25\,h & 53\,h & 19\,h \\
\hline
\end{tabularx}
\caption{Simulation time $t_\text{3D}$ for one cardiac cycle of the 3D simulation, model time constant $\bar{\tau}/T$, and number of cardiac cycles (pressure / flow) for different initial conditions (zero, steady, 1D) with $\epsilon_\infty\leq1\,\%$. The maximum number of cardiac cycles in a simulation is highlighted in bold. The last row shows the time savings on 48 CPU cores for each simulation when using the 1D initialization compared to the steady state initialization.\label{tab_models_cycles}}
\end{table}

\section{Discussion}

% Alison: expanding more on the ideas around how easy it is to make a mistake in thinking the simulation is converged when using standard methods (even add some examples of cycle to cycle variation?)   
We gave a detailed review of properties of lumped-parameter boundary conditions in cardiovascular fluid dynamics simulations. The speed of convergence to a periodic state solution is determined by a single model time constant, that can be approximated as the mean of all individual time constants of multiple outlet boundary conditions. We found that for 53 models of the Vascular Model Repository, the model time constant $\bar{\tau}/T$ spanned from 0.3 in pulmonary models to 9.6 in aorta models. That means that in the most extreme case, the time constant of the model is almost ten times as large as the length of the cardiac cycle. Using zero initial conditions, the number of cardiac cycles required to reach a periodic state scales linearly with the model time constant, reaching a median of eight cardiac cycles for our subset of models. Even when using initial conditions other than zero, e.g., from a steady state solution, the number of cardiac cycles required still scales with model time constant.

\begin{figure}[hbt!]
\centering
\includegraphics[height=5.5cm]{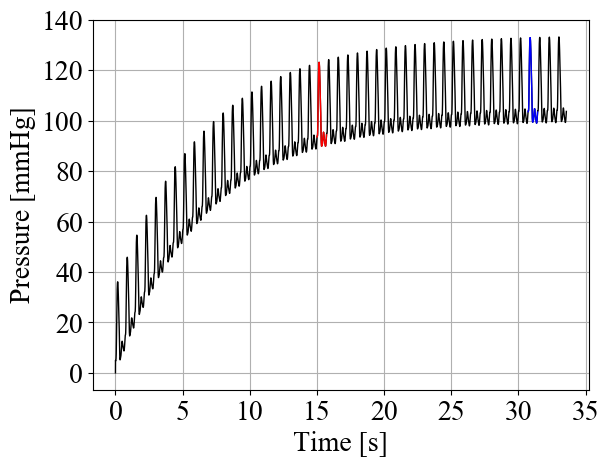}\hspace{1cm}
\includegraphics[height=5.5cm]{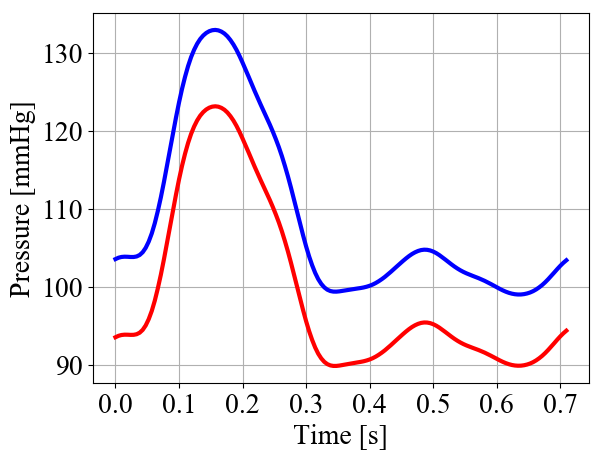}
\caption{Pressure convergence of an outlet in model 0076\_1001 with time constant $\bar{\tau}/T=9.6$, starting from zero initial conditions. The red cardiac cycle was selected at $\epsilon_n=1\,\%$, the blue one at $\epsilon_\infty=1\,\%$. \label{fig_asymp_errors}}
\end{figure}

We introduced the asymptotic error to determine whether a simulation has reached a periodic state, calculating the relative difference of the results in the current cardiac cycle to a perfectly periodic solution. As the periodic solution is not known \emph{a priori}, we proposed to use a 0D model of the boundary condition. Taking the flow at a 3D outlet as input, we estimate the periodic pressure and compare it to the pressure at the 3D outlet. This method is a quick and easy validation tool for any 3D vascular fluid dynamics simulation with lumped-parameter boundary conditions. In models with time constants larger than $1/\ln 2 \approx 1.44$, the cycle-to-cycle difference is lower than the asymptotic error. A low cycle-to-cycle difference can erroneously lead the user to believe that the simulation has already reached a periodic state when comparing two consecutive cardiac cycles. This is visualized in Figure~\ref{fig_asymp_errors}. Using the same error tolerance of 1\,\% but different error norms, cyclic error (red) and asymptotic error (blue), yields a pressure difference of 10\,mmHg. Thus, only the asymptotic error is a suitable metric to determine the distance of a simulation to its periodic state.

To reduce 3D computation times, we proposed a method to generate good initial conditions in an automated and computationally inexpensive way. We used SimVascular to automatically generate a 1D replication of the 3D model, which we ran until we achieved a periodic state. We than mapped the 1D solution onto the 3D Finite Element mesh.  Note that the mapping does in general not result in a fluid field that is divergence free. However, we are not interested in generating a physical or physiological meaningful solution. Instead, the initial conditions are ``washed out" by the first iteration of our numerical solver of the Navier Stokes equations. Similarly, the pressure field does not capture local variations. Here, it should be noted that it is much more important to match the overall pressure level of the model, which is in general much higher than any local variations. In a convergence analysis with six different vascular models, we demonstrated that models with our 1D initialization method converge in one or two cardiac cycles. This greatly reduces the computation time for the 3D model over the standard method of initializing pulsatile simulations with a steady-state solution, typically dozens of hours in simulation time and hundreds of hours in CPU time. The code for the 1D initialization is freely available on GitHub (\url{github.com/SimVascular/SimVascular}).

We close with a discussion of the the limitations and future perspectives of our work. We did only consider RCR boundary conditions in this work. However, there are many more examples of lumped parameter 0D networks that are coupled to 3D models, such as coronary boundary conditions \cite{kim10a,kim10b} and 0D closed loop models used, e.g., in simulations for single ventricle \cite{bove07,bove08} and coronary artery bypass graft (CABG) patients \cite{ramachandra16}. In the general case of lumped parameter boundary conditions, the model time constant cannot be determined analytically. However, it can still be calculated numerically from the time it takes the boundary condition to respond to a step in the inflow condition. We plan to include these boundary conditions in future work. It should be noted that those boundary conditions not only require initialization of the velocity and pressure field in the 3D model but also in unknowns that are internal to the 0D model. Furthermore, we only considered rigid-wall simulation in this work. We plan to test our 1D initialization method for deformable wall simulations as well, using the coupled momentum method \cite{figueroa06} or an arbitrary Lagrangian–Eulerian formulation \cite{baeumler20}. To further improve our method and guarantee periodic convergence within one cardiac cycle, it will be necessary to improve the mapping of the velocity field to the 3D model. This could be accomplished for example by solving a Stokes flow problem, using the 1D results as boundary conditions.

\section{Acknowledgments}
We thank Dr. Luca Pegolotti and Dr. Irene Vignon-Clementel for insightful discussions. This work was supported by NIH grants R01LM013120 and R01EB029362. The authors gratefully acknowledge the Stanford Research Computing Center for providing the computational resources necessary to the numerical simulations presented in this work.

\bibliographystyle{unsrt}
\small
\bibliography{references}

\end{document}